\documentclass{sig-alternate-10pt}

\usepackage{graphicx}

\usepackage{fixltx2e}

\usepackage{url}

\usepackage{xcolor}
\usepackage{amssymb}
\usepackage{calc}
\usepackage{booktabs}

\usepackage{float}

\usepackage{microtype}
\usepackage{mdwlist}
\usepackage{lipsum}
\usepackage{blindtext}
\usepackage[english]{babel}

\usepackage[caption=false]{subfig}

\usepackage{clrscode}
\usepackage{algorithmic}
\usepackage{algorithm}
\definecolor{algComment}{rgb}{0.4,0.4,0.4}

\usepackage[draft,footnote,nomargin]{fixme}
\usepackage{todonotes}
\newcommand{\refFig}[1]{Figure~\ref{#1}}
\newcommand{\refSec}[1]{Section~ \ref{#1}}

\newcommand{\refTab}[1]{Table~\ref{#1}}

\newcommand{\refAlg}[1]{Algorithm~\ref{#1}}

\definecolor{blind-gray}{gray}{0.9}

\newcommand{\tih}[1]{}
\begin{document}

% \conferenceinfo{CoNEXT'12,} {December 10--13, 2012, Nice, France.}
% \CopyrightYear{2012}
% \crdata{978-1-4503-1775-7/12/12}
\clubpenalty=10000
\widowpenalty = 10000

\title{Anticipatory Buffer Control and Quality Selection for Wireless Video Streaming}
%\title{SmarterPhones: Anticipatory Buffer Control and Quality Selection for Wireless Video Streaming}
%\subtitle{}

\numberofauthors{1} %  in this sample file, there are a *total*
\author{
\alignauthor
Martin Dr\"axler$^{\dag}$, Johannes Blobel$^{\dag}$, Philipp Dreimann$^{\dag}$, Stefan Valentin$^{\ddag}$, Holger Karl$^{\dag}$\\
\affaddr{$^{\dag}$University of Paderborn, Germany} \affaddr{$^{\ddag}$Bell Labs, Alcatel Lucent Stattgart, Germany}\\
\email{\{martin.draexler, johannes.blobel, philipp.dreimann, holger.karl\}@upb.de}\\
\email{stefan.valentin@alcatel-lucent.com}
}
% There's nothing stopping you putting the seventh, eighth, etc.
% author on the opening page (as the 'third row') but we ask,
% for aesthetic reasons that you place these 'additional authors'
% in the \additional authors block, viz.
% \additionalauthors{Additional authors: John Smith (The Th{\o}rv{\"a}ld Group,
% email: {\texttt{jsmith@affiliation.org}}) and Julius P.~Kumquat
% (The Kumquat Consortium, email: {\texttt{jpkumquat@consortium.net}}).}
\date{\today}
% Just remember to make sure that the TOTAL number of authors
% is the number that will appear on the first page PLUS the
% number that will appear in the \additionalauthors section.

\maketitle

%\blfootnote{This work was partly supported by Bell Labs, Stuttgart within the research collaboration Smarter Phones And smarter Networks (SPAN).}

\begin{abstract}
Video streaming is in high demand by mobile users, as recent studies indicate. In cellular networks, however, the unreliable wireless channel leads to two major problems. Poor channel states degrade video quality and interrupt the playback when a user cannot sufficiently fill its local playout buffer: \emph{buffer underruns} occur. In contrast to that, good channel conditions cause common greedy buffering schemes to pile up very long buffers. Such \emph{over-buffering} wastes expensive wireless channel capacity.

To keep buffering in balance, we employ a novel approach. Assuming that we can predict data rates, we plan the quality and download time of the video segments ahead. This \emph{anticipatory scheduling} avoids buffer underruns by downloading a large number of segments before a channel outage occurs, without wasting wireless capacity by excessive buffering. We formalize this approach as an optimization problem and derive practical heuristics for segmented video streaming protocols (e.g., HLS or MPEG DASH). Simulation results and testbed measurements show that our solution essentially eliminates playback interruptions without significantly decreasing video quality.

% when the user decides to jump inside a video or to watch another.

%Video delivery solutions often use streaming protocols, e.g. HTTP Live Streaming (HLS) or MPEG DASH, that download the video stream in segments. This segmentation enables two degrees of freedom: \emph{when} to download segments and potentially puffer them, and in \emph{which quality} to download segments.
%
%By combining these two decisions with anticipatory knowledge of a mobile user's future data rate, we can create a download schedule.
% %that indicates \emph{when} to download which segment at \emph{which quality}.
%This schedule can minimize playback interruptions and maximize the video quality. Our approach can both improve the QoE for the users and enable a more efficient use of wireless network recourses.

%For this approach we formulate both an optimization problem and heuristic algorithms. We evaluate all solutions by both simulations and by a testbed implementation. On the basis of the testbed we can also show how our approach can be integrated into existing systems.
%
%With the results for our scenarios we can show that our approach essentially eliminates playback interruptions without significantly decreasing video quality.

\end{abstract}

\category{C.2.1}{Computer-Communication Networks}{Network Architecture and Design}[Wireless communication]
\category{F.2.2}{Analysis of Algorithms and Problem Complexity}{Nonnumerical Algorithms and Problems}[Sequencing and scheduling]
\category{H.5.1}{Information Interfaces and Presentation}{Multimedia Information Systems}[Video]
\terms{Algorithms, Design, Performance, Measurement}
\keywords{Video Streaming, HLS, MPEG-DASH, Scheduling}

%A category including the fourth, optional field follows...
%\category{D.2.8}{Software Engineering}{Metrics}[complexity measures, performance measures]

%\keywords{ACM proceedings, \LaTeX, text tagging} % NOT required for Proceedings

%%% OLD STRUCTURE CONSIDERATIONS %%%

%\ti{based on \cite{1309.5491}}

\section{Introduction}\label{sec:intro}
%1 P -- SV
Delivery of video content over wireless broadband networks is already widely used today and is expected to increase heavily in the upcoming years. Studies by Cisco \cite{CiscoVNI} and
Akamai \cite{Akamai} indicate that mobile data traffic will increase by a factor of 25 from 2011 to 2016 with around two-thirds of this traffic being streamed video traffic. The wireless
infrastructure cannot keep up with this trend by merely increasing data rate. It is necessary to organize mobile data transmission in a better way, as also indicated by Akamai \cite{Akamai}.

We present an approach to combine buffer control and video quality selection based on anticipation of wireless data rates. Our approach and the following motivation
is based on the HTTP Live Streaming (HLS) protocol \cite{hls,hlsapple}, but can also be applied to similar video streaming protocols, like MPEG DASH \cite{lederer2012dynamic,6077864}.

In HLS
a video is not transmitted as a continuous stream of data, but it is divided into \emph{segments} of a certain duration and then transmitted segment by segment. These segments are downloaded
via HTTP from the server and are then concatenated by the player application for playback. For example, a video of $120$\,s and segments of $10$\,s would be divided into 12 segments. This implies
that for uninterrupted playback, segment $i+1$ has to be downloaded before segment $i$ has been played to its end in the HLS player application. If a segment is downloaded before it is needed for playback,
it is buffered at the HLS player application.

Another feature of the HLS protocol is video quality selection: each segment can be present on the server in different quality levels. A quality level is determined by the resolution and/or the encoding
bit rate of the video and is then identified in HLS by the resulting file size of the video segment. As our approach optimizes downloading of video segments and the file
size has a direct implication on the required data rate for a download, we adopt this definition of video quality for this paper. Nonetheless, this definition of quality should not be confused with visual video quality metrics like PSNR or MOS or the pure bit rate of a video codec.

To download a segment, the player application has to decide in which quality level to download it. This is done in current
HLS-compatible players like VLC \cite{VLC} or the Apple iOS and Android media players, but the selection only relies on the measurement of the current and past data rates.
In order to integrate anticipatory knowledge of future data rates into our approach we use what we call \emph{channel anticipation}. The idea behind this is very similar to classical
channel prediction used for improved scheduling decisions in mobile access networks, but our approach works on different time scales and accuracy levels. The time scales in which our approach has to work are defined by the length of the video segments, which are usually on the order of tens of seconds, in contrast to channel prediction for a few milliseconds. Furthermore, we are interested in rough estimates of achievable data rates and not precise channel quality samples. We describe this idea in further detail in \refSec{sec:anticipation}.

With this idea we extend the default behavior in the HLS protocol by explicit buffer control and quality selection based on the anticipated data rates. The motivation for this extension is
straightforward: As long as enough data rate is
available in the future, the HLS video player should not download and buffer too many segments. Buffering too many segments in this case has no benefit for the user's QoE, but may have the downside of
using wireless resources that could otherwise be used to benefit other users. We call this problem \emph{over-buffering}. If there is a future decrease in available data rate, the HLS player has to
download and buffer more segments in advance. If the HLS player does not download enough segments in advance the playback will stall; we call this problem \emph{buffer underrun}. In parallel to this decision on \emph{when} to
download segments is the decision in \emph{which quality} to download segments. If the data rate is insufficient to download segments in a high quality, but a lower quality is available, the
HLS player should switch to the lower quality to prevent a buffer underrun.

We call this combination of \emph{when} to download each segment in \emph{which quality} a schedule. Hence, such a schedule is only executed on the application layer. For the physical layer schedule we assume that a normal, fair scheduler has already assigned radio resources to the users. This makes our scheduling independent from the physical layer scheduling of different wireless technologies. Additionally this allows us to perform our scheduling for each user individually, because the physical resources are already shared and we do not have to consider any resource sharing.
Hence, the anticipated wireless data rates are actually achievable and effects like number of users per cell are already incorporated by the anticipation scheme.

In \refSec{sec:optimization}, we present an optimization problem to create such a schedule and
introduce heuristic algorithms to compute schedules in \refSec{sec:heuristics}. There we also illustrate examples for the described scheduling decisions. In \refSec{sec:system}, we explain how our scheduling approach
can be integrated into an existing system and describe how we developed a testbed implementation. We use this testbed implementation together with a simulation in \refSec{sec:eval} to evaluate
our approach and to present the results. We conclude our work in \refSec{sec:concl}. Before presenting the optimization problem we first give an overview of existing related work in the next
section.% and discuss channel anticipation in \refSec{sec:anticipation}.

%\ti{continue: anticipation, over/under-buffering, example, paper structure}
 
% Should motivate/explain:
% \begin{itemize}
% \item Playback interruptions
% \item Video quality (or a better name for that)
% \item Overbuffering (network operator) , Underbuffering (user)
% \end{itemize}
% 
% \bt\bt\bt\bt

\section{Related Work}\label{sec:relwork}
% \cite{ProebsterKWV12}
% \cite{1580942,LuTR}
% \cite{6364845,6364725}
% \cite{Riiser,Muller}
%\ti{section intro, fix intro to new structure}

In this section we first give an overview of existing work on adaptive video streaming and then explain how existing mechanisms can be used to implement an approach for channel anticipation.

\subsection{Adaptive Video Streaming}
\label{sec:adaptivevideostreaming}

There is a large body of work on techniques for adaptive video streaming. At the application layer, various control loops adapt video quality \cite{Jiang:2012:IFE:2413176.2413189,Tian:2012:TAS:2413176.2413190,Riiser,Muller} to the end-to-end data rate and channel-aware pre-fetching invokes a traffic burst at high channel gain \cite{shih-fu:05,reisslein97:vbr_prefetching}. At the link layer, cross-layer schedulers to jointly adapt video quality and wireless resource allocation have been proposed \cite{huang:08,1580942,lu13:anticipative_video_delivery}. 

Compared to this work our approach differs twofold. First, our approach does not adapt video quality or wireless resources alone. Instead, it joins video quality adaptation with the allocation of playout buffer size. Unlike \cite{huang:08,1580942}, this enables to trade off video quality against the amount of allocated resources.

Second, our adaptation is \emph{anticipative} and not reactive. Unlike any of the above approaches, our joint buffer-quality allocation is based on a prediction of the user's wireless data rate. Using this prediction enables our scheme to plan ahead, when to download a video segment at which quality. This enables to compensate for upcoming channel outages (e.g., when a user drives through a tunnel) by downloading segments in advance. Although this idea of anticipation has been applied for software interfaces \cite{nadin00:anticipatory} and cognitive radios \cite{tadrous11:proactive_ra}, it has not been proposed for media streaming so far. 

Further benefits of our work are its generality and practical computational complexity. Our heuristics run on general-purpose hardware at high speed and their formulation is entirely based on bit rates. This captures arbitrary variable bit rate protocols for audio and video streaming without using subjective quality metrics. This level of tractability is not provided by Utility-based formulations such as \cite{huang:08}.
%\ti{HK: warum?}

\subsection{Channel Anticipation}
\label{sec:anticipation}
To compensate for upcoming channel outages, the future state of the wireless channel has to be estimated. We assume this \emph{channel anticipation} to operate on the order of seconds, which is three orders of magnitude above typical channel predictors \cite{min07,schmidt11,akl12:fb_delay_opt}. At such a large time scale, error control and resource allocation already have compensated for and averaged out fast fading leaving user mobility as dominating cause of channel state variation. As a user moves through the cellular network, its path loss towards the base stations becomes time-variant causing interference and channel gain to vary at a large time scale. Only these large-scale dynamics need to be accounted for when anticipating the user's data rate for tens of seconds in advance.

This anticipation cannot be solely based on PHY measurements such as Channel Quality Indicators (CQIs) but needs to include information about the user's environment. Such context information usually comes in two forms, either as a database based on past data collection, or as live information from handsets and base stations. In particular, we consider the databases
\begin{itemize*}%[noitemsep,nolistsep]
\item Coverage and capacity maps
\item Load and interference maps
\item Maps of streets and similar features for land navigation (tunnels, railroads, points-of-interests)
\item Common patterns of user behavior (e.g. trajectories, speed, bearing)
\end{itemize*}
and the live input
\begin{itemize*}
\item Channel state, e.g., from handset
\item Load and interference, e.g., from base stations
\item Localization information, e.g., from GPS, handset sensors, and cellular network
\item Current and planned trajectory, e.g., from turn-by-turn navigation
\end{itemize*}
%as available information for our channel anticipation mechanism.
as available information for our anticipation mechanism.

Based on this information, various methods to anticipate the data rate of wireless devices have been presented. Focusing on users with homogeneous trajectories, Riiser et al. presented accurate predictors in \cite{Riiser2012} and a prototype in \cite{Riiser2013}. Based on this work for users in a bus, train, ferry or car, Yao et al.\ studied a similar approach for users in public transport \cite{Yao2011} and cars \cite{Yao2008,Yao2012}. The resulting maps have been used by Fardous et al.\ \cite{Fardous2012} to build an anticipation mechanism that incorporates both live locations and planned trajectories from the car's turn-by-turn navigation system. Likewise, future work may use the trajectories pre-computed by autonomous vehicles \cite{leonard08} to anticipate wireless data rate.

%Independent of a specific means of transport, the authors of \cite{Sridharan2013,Isaacman2012,Noulas2012} have characterized regular movement patterns for users of mobile phones. Such regularity has been exploited with the help of Markov chains \cite{Chen2013} or Support Vector Machines \cite{Anagnostopoulos2012,chen13:localization_ml} to predict the user's trajectory through a cellular network. Along those future positions, lookups in the above capacity maps and interpolation can provide the user's data rate.

These studies consistently conclude that data rates of homogeneously moving users can be accurately anticipated for tens of seconds ahead. Further studies point to a strong spatial and temporal regularity of the user's trajectories \cite{Sridharan2013,Isaacman2012,Noulas2012}, which allows to accurately predict its position for several seconds in advance \cite{Anagnostopoulos2012,chen13:localization_ml}.

Based on this evidence, we assume accurate rate anticipation for users on highways, in trains, or public transportation as a basis of this paper.
By limiting our focus to users with homogeneous trajectories we do not only make reasonable assumptions on rate anticipation. Moreover, we target scenarios where users naturally have a high demand for wireless video streaming \cite{sandvine13}.

\section{Optimization Problem}
\label{sec:optimization}
%2 P -- MD
The previously introduced scheduling problem can be formulated as a mixed-integer, quadratically constrained (MIQCP) optimization problem. We have presented a simpler version in \cite{draexler13:span}; here we extend this approach by also incorporating the buffer fill level into the constraints and objective of the MIQCP.

\subsection{Assumptions}
We assume a discrete time model. Time is represented as a sequence of time slots $t_i$ of constant length. For simplification, we further
assume that the length of each time slot is equal to the playback duration of one video segment. Thus time slots and segments are unitless and can be used together in a constraint.
Additionally, each video segment has to be downloaded within exactly one
time slot, i.e. the download of a video segment must not be spread across multiple time slots. This implies that for an uninterrupted playback of a video, the $i$-th video
segment has to be downloaded within time slot $t_i$ or earlier.
Downloads in a given time slot are limited by the data rate for each user in this slot. Each user is connected to at
most one base station per $t_i$. We assume that the allocation of data rates to the users is done by an underlying, non-modifiable radio resource scheduler, limiting our scheduling approach for the download of
video segments to a higher layer.
The file size for each video segment, i.e the required amount of data to download, is determined by the selected video quality level. The data rate limits and the video quality levels are given
in the same units.

\subsection{Formulation}
%OP from IWMC paper, briefly
The optimization problem takes the parameters listed in \refTab{tab:opparams} as input.% The parameters correspond to the previously described assumptions for modeling the optimization problem.
%In principle, two decisions have to be made in order to solve the anticipatory scheduling problem.
In principle, two decisions have to be made in order to solve the scheduling problem.

First, for each video segment $s$ the time slot $t$ in which to download the segment
has to be determined. This decision has to be taken for each user $u$ and is represented by the variable $d_{s,u}$:
\begin{align}
%d_{s,u} = t, t \in T, \forall s \in S, u \in U
d_{s,u} \in T, ~\forall s \in S, u \in U
\end{align}

Second, for each video segment $s$ one video quality level $q$ from the set of available qualities $Q$ has to be selected as variable $p_{s,u}$ for each user $u$, assuming that all segments are available in all qualities:
\begin{align}
%p_{s,u} = q, q \in Q, \forall s \in S, u \in U
p_{s,u} \in Q, ~\forall s \in S, u \in U
\end{align}

%The amount of downloadable segments per time slot $t$ is constrained by the data rate for each user in the time slot:
The amount of downloadable segments per time slot $t$ is limited by the data rate for each user in the time slot:
\begin{align}
\sum_{d_{s,u} = t} p_{s,u} &< C_{u,t}, u \in U, ~\forall s \in S, \forall t \in T%\\
%\sum_{d_{s,u} = t} p_{s,u} &< C_{t}, \forall s \in S, \forall u \in U, \forall t \in T
\end{align}
As there is no resource sharing among users on the application layer, it is sufficient to consider each user separately.

\begin{table}[tb]
\begin{center}
\caption{MIQCP input parameters}
\begin{tabular}{ll}
\toprule
$T$ & set of time slots, $t \in \mathbb{N}$\\
$S$ & set of segments to transfer, $s \in \mathbb{N}$\\
$U$ & set of users\\
%$C_t$ & overall BS capacity at time $t$, $C_t \in \mathbb{Q}^{+}$\\
%$C_{u,t}$ & individual channel capacity for user $u$\\&at time $t$, $C_{u,t} \in \mathbb{Q}^{+}$ \\
$C_{u,t}$ & data rate for user $u$ at time $t$, $C_{u,t} \in \mathbb{Q}^{+}$ \\
$Q$ & set of segment video quality levels, $q \in \mathbb{Q}^{+}$\\% e.g. $\{10,20,50\}$
\bottomrule
\end{tabular} 
\label{tab:opparams}
\end{center}
\end{table}

%To include belated downloads of HLS video segments into the objective function, we need to calculate the lateness of each segment $s$ as a variable $l_{s,u}$ for each user $u$. The lateness of a segment
To include belated downloads of video segments into the objective function, we need to calculate the lateness of each segment $s$ as a variable $l_{s,u}$ for each user $u$. The lateness of a segment
should be $0$ if it is downloaded in time, irrespective of how early it was downloaded:
\begin{align}
l_{s,u} = \mathrm{max}(d_{s,u}-s,0), ~\forall s \in S, u \in U
\end{align}

We also want to take the number of \emph{buffered} segments for each user $u$ in time slot $t$ into account and calculate it as variable $b_{t,u}$ by summing up the number of downloaded segments
until $t$ and subtracting $t$ (because up to time slot $t$, $t$ segments had to be played out already):
\begin{align}
%b_{t,u} = \sum_{d_{s,u} \leq t \forall s \in S} - t, \forall t \in T, u \in U
%b_{t,u} = | d_{s,u} \leq t \forall s \in S | - t, \forall t \in T, u \in U
%b_{t,u} =  \mathrm{min}\left(\left(\sum_{\forall s \in S~ d_{s,u} \leq t} 1 \right) - t,0 \right),~\forall t \in T, u \in U
b_{t,u} =  \mathrm{min}\left(\left(\sum_{\substack{\forall s \in S\\ d_{s,u} \leq t}} 1 \right) - t,0 \right),~\forall t \in T, u \in U
\end{align}

To formulate the objective function we define three weight factors: $W_l$ for the lateness of video segments, $W_q$ for the selected quality level of the video segments and $W_b$ for the number
of buffered segments. Now we can now formulate the objective function as
\begin{align}%\nonumber \\
\text{minimize: } W_l \ \cdot \sum_{s \in S, u \in U} l_{s,u} &- W_q \cdot \sum_{s \in S, u \in U} p_{s,u} \nonumber \\ &+ W_b \cdot \sum_{t \in T, u \in U} b_{t,u}
\end{align}
%This formulation allows both to trade-off the metrics lateness, video quality, and buffer usage as well as to define a lexicographical ordering of these metrics, which we will do for  our evaluation.
This formulation allows to define a lexicographical ordering of the metrics lateness, video quality, and buffer usage, which we will do for our evaluation.
As an alternative to this objective function it is also possible to set fixed limits to one or two of the metrics and to maximize or minimize the remaining ones, deriving a corresponding Pareto front.
%\ti{CoNEXT Review: Explaining the objective function in some detail rather than simply stating it in eq. 6 would improve the readability of the paper.}

The described formulation of the optimization problem is not positive semidefinite, but a very compact and straightforward formulation. We described a positive semidefinite formulation in \cite{draexler13:span}, which can
directly be solved using a standard solver for mixed integer quadratically constrained optimization problems.

\section{Heuristics}
\label{sec:heuristics}
We analyze two different types of heuristic algorithms for our scheduling problem: two greedy scheduling algorithms, which illustrate the behavior of standard HLS player applications, and
the \proc{Fill} algorithm, which tries to minimize playback interruptions while keeping buffering minimal.

% Both the greedy algorithms and the \proc{Fill} algorithm only solve the scheduling of segment downloads and the video quality selection, and not the allocation of the available transmission
% capacity among the users. Thus the allocated capacity for each user in each time slot is calculated by dividing the overall capacity for a base station by the number of users associated to this
% base station in the given time slot.

Consistent with the optimization problem, all heuristics are offline schedulers, which means the data rates for all time slots are known in advance and the result of all heuristics is a complete schedule for all users over a given number of time slots. The assumptions are the same as for the optimization problem.

\subsection{Greedy Scheduling}
\label{sec:greedyscheduling}

Both greedy scheduling algorithms take the available data rate for each user in each time slot and a maximum buffer size as their input. Based on their respective objective, they iterate over
all time slots and decide which segments to download to fill the buffer with video segments. In each time slot, they consider the currently available data rate, the current number of buffered segments and the quality levels of the segments not yet scheduled.

The greedy algorithms cannot adapt the maximum buffer size, which can result in unnecessary buffering if enough data rate is available to play the video without buffering, or unwanted playback interruptions if the chosen buffer size is not big enough to continue playback in phases of insufficient data rate.

\subsubsection{BufferFirst Algorithm}

The objective of the \proc{BufferFirst} algorithm is to fill the entire buffer with video segments. If the buffer is not completely full in a time slot the algorithm schedules the download of the maximum possible amount of segments at the lowest quality supported by the currently available data rate and free buffer space. If there is also enough data rate available to download segments in higher quality levels, it increases the quality for the downloaded segments.
Thus, the algorithm will never decide to download fewer segments to increase the quality.
%It selects the quality of segments such that the buffer is filled. If there is enough data rate available the algorithm selects segments with higher quality, but never decides to download fewer segments to incrase the quality.
%It selects the highest quality with which the buffer
%can be filled.

\begin{figure}[htb]
  \begin{center}
  \includegraphics[width=0.315\textwidth]{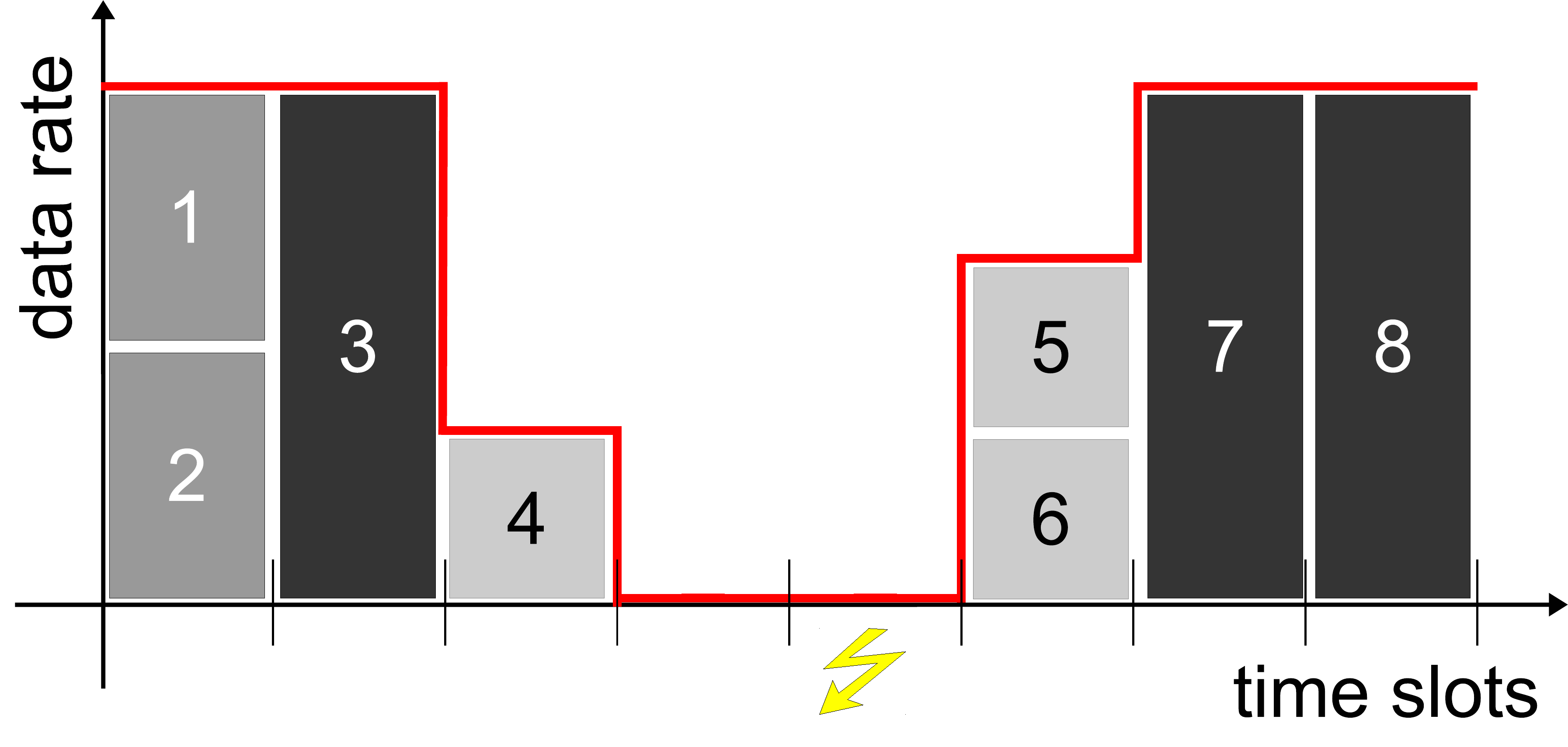}
  \caption{Example for BufferFirst Algorithm}
  \label{fig:scheduler_bufferfirst}
  \end{center}
\end{figure}

\refFig{fig:scheduler_bufferfirst} shows an example for the \proc{BufferFirst}
algorithm with a maximum buffer size of two segments. The rectangles show the video segments with different qualities (indicated by their shading) and the solid line above the rectangles indicates
the available data rate. The buffer is filled with segments of medium quality in
the first time slot; in the second and third time slots, one segment is
downloaded to fill the buffer again. With this schedule the video playback will
not be interrupted in the fourth time slot,
%because there is still one segment in the buffer, 
but the video playback is interrupted in the fifth time slot.

\subsubsection{QualityFirst Algorithm}

The objective of the \proc{QualityFirst} algorithm is to download segments with the highest quality possible. If the buffer is not completely full in a time slot the algorithm schedules the download of new segments at the maximum possible quality supported by the currently available data rate. If there is still free buffer space and data rate it continues to schedule downloads of further segments.

%The \proc{QualityFirst} algorithm first tries to download segments with the highest available video quality level if there is enough data rate and buffer space left. If there is not enough data rate
%left, it tries to download segments with the next lower video quality level, until there is no data rate or buffer space left.
As a consequence, this algorithm favors downloading segments at higher video quality levels at the expense of buffering segments.

\begin{figure}[htb]
  \begin{center}
  \includegraphics[width=0.315\textwidth]{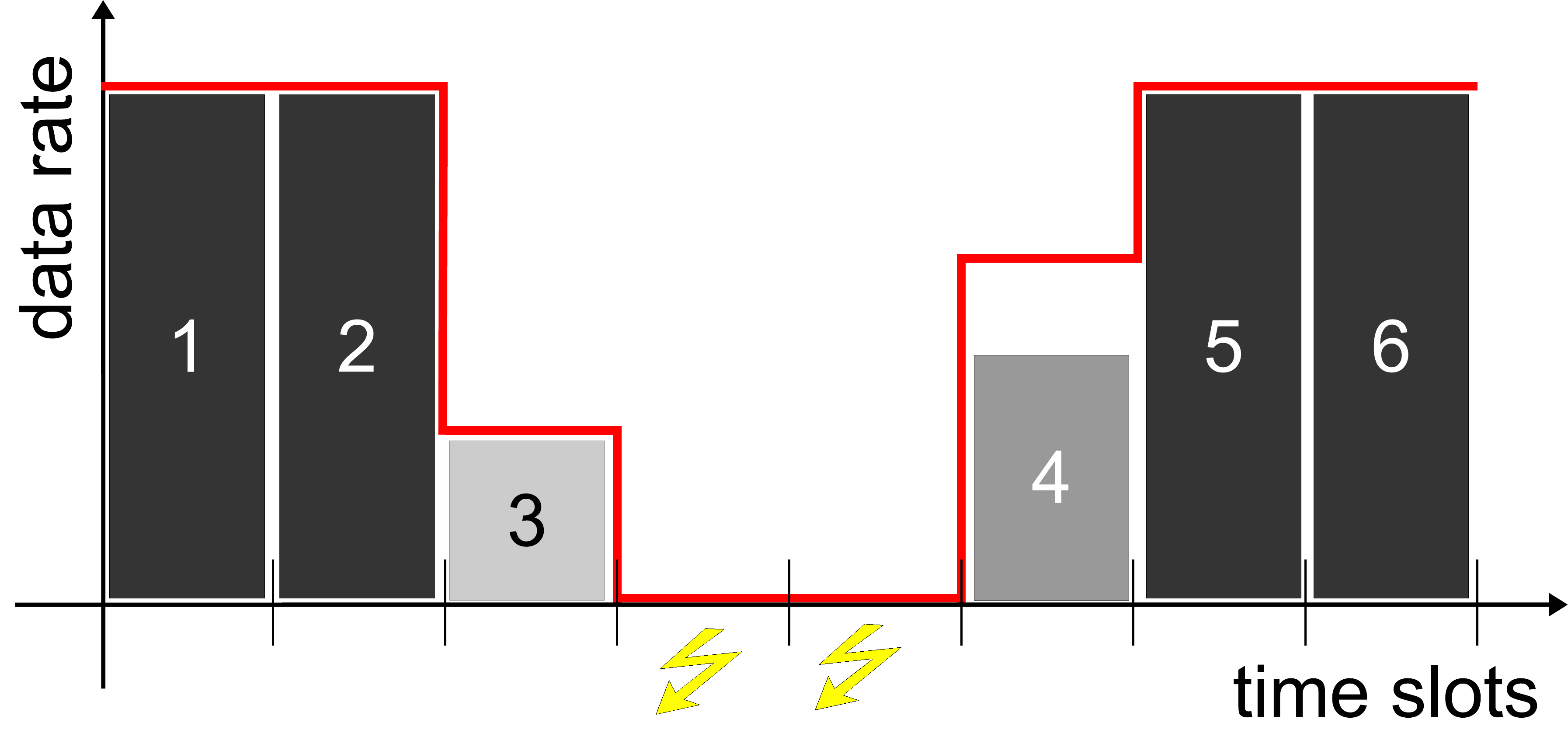}
  \caption{Example for QualityFirst Algorithm}
  \label{fig:scheduler_qualityfirst}
  \end{center}
\end{figure}

\refFig{fig:scheduler_qualityfirst} shows an example for the \proc{QualityFirst} algorithm with a maximum buffer size of two segments. As the algorithm favors to download segments with high quality levels
before filling the buffer, there are no segments in the buffer to avoid a playback interruption in the fourth and fifth time slot.

\subsection{Fill Scheduler}
%\section{Fill Scheduler Draft}
\label{sec:fillscheduler}
\tih{MD: hier habe ich einige Dinge umgebaut, das Layout passt noch nicht, sodass die Algorithms und Figures nicht unbedingt da sind, wo sie idealerweise sein sollten.}

We designed the \proc{Fill} algorithm to eliminate the shortcomings of a fixed buffer size.
%\todo[inline]{Wo? -- Bei Section~\ref{sec:greedyscheduling} wird das zwar erwähnt, ist aber IMHO nicht ausreichend weil dieses Problem nicht in den Beispielen erläutert wird. Beim BufferFirst würde sich dies anbieten.}
Identical to the greedy scheduling algorithms, the \proc{Fill} algorithm takes the available data rate for each user in each time slot as its parameter and it also iterates over all time slots to fill the buffer with video segments (independently for all users).
%The algorithm calculates download schedules for all users independently and iterates over all time slots for each user as shown in \refAlg{alg:fillscheduler}. 
\refAlg{alg:fillscheduler} shows this main structure.
%The function $\proc{calculateUserRates}(u)$ calculates available data rates for a user for all time slots
The function $\proc{anticipateUserRates}(u)$ returns anticipated data rates for a user for all time slots 
%by taking the minumum of the user's inidvidual
%channel capacity and the allocated fraction of the overall base station capacity.
based on the underlying radio resource scheduler and channel anticipation.

\begin{figure}[tb]
    \centering
    \includegraphics[width=0.475\textwidth]{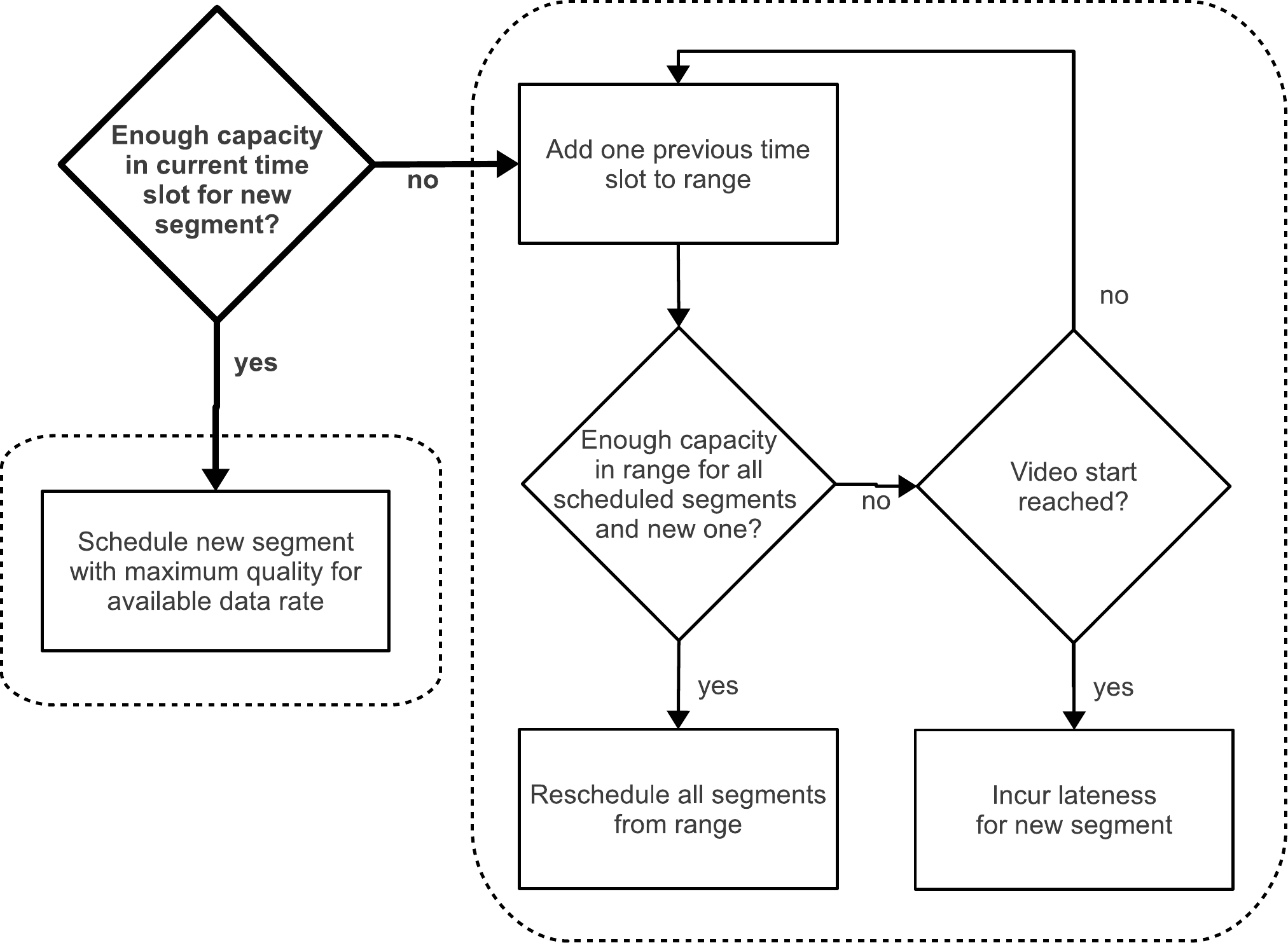}
    \caption{Flowchart for Fill Scheduler}
    \label{fig:fillflowchart}
\end{figure}

\begin{algorithm}[!ht]
%\scriptsize
\small
\caption{$\proc{fillScheduler}(U,T,Q)$}
\label{alg:fillscheduler}
\algsetup{indent=0.5em}
\begin{algorithmic}[1]
\STATE \COMMENT{users U, times T, qualities Q}
\FORALL[schedule all users]{$u \in U$}
\STATE $C \gets \proc{anticipateUserRates}(u)$ \COMMENT{from channel anticipation}
\STATE $s \gets 0$ \COMMENT{initialize couter for scheduled segments}
\FORALL[schedule all timeslots/segments]{$t \in [0..|T|]$}
\STATE $s \gets s + \proc{scheduleSegment}(u,t,s,Q,C)$
\ENDFOR
\ENDFOR
\end{algorithmic}
\end{algorithm}

The basic operation of $\proc{scheduleSegment}$ is illustrated in \refFig{fig:fillflowchart}. For each time slot there are two different operations possible, depending on the available data rate in the time slot.

\begin{algorithm}[!hbt]
\tih{MD: kommentare im algorithm passend zum flowchart}
%\scriptsize
\small
%\footnotesize
\caption{$\proc{scheduleSegment}(u,t,s,Q,C)$}
\label{alg:fillschedulesegment}
\algsetup{indent=0.5em}
\begin{algorithmic}[1]
\STATE $q \gets \proc{getBestQuality}(Q,C[t])$
\IF[enough capacity in current time slot for new segment?]{$q \neq \FALSE$}

\STATE $\proc{schedule}(u,s,t,q)$ \COMMENT{schedule new segment with maximum quality for available data rate}
\RETURN $1$

\ELSE[even lowest quality not feasible in time slot $t$]

\FORALL{$g \in [t .. 0]$}
\STATE \COMMENT{enough capacity in range $[g..t]$ for all scheduled segments and new one?}
% \IF{$\proc{getBestQualityRange}(Q,t-g+1,\sum\limits_{i=g}^t C[i]) \neq \FALSE$}
% \STATE \COMMENT{going back to $g$ provides enough data rate}
\IF[going back to $g$ provides enough data rate]{$\proc{getBestQualityRange}(Q,t-g+1,\sum\limits_{i=g}^t C[i]) \neq \FALSE$}
\STATE $q \gets \proc{getBestQualityRange}(Q,t-g+1,C[g...t])$
\STATE $p \gets 0$
\FORALL[reschedule all segments from range]{$r \in [g .. t]$}
\STATE $n \gets \proc{getSegmentsForQuality}(q,C[r])$

\FORALL{$v \in [(g+p) .. (g+p+n)]$}
\STATE $\proc{schedule}(u,v,r,q)$
\ENDFOR

\STATE $p \gets p + n$
\ENDFOR
\RETURN $1$
\ENDIF
\ENDFOR ~ \COMMENT{video start reached}
\RETURN $0$ \COMMENT{incur lateness for new segment}

\ENDIF
\end{algorithmic}
\end{algorithm}

If there is enough data rate to download a new segment in the currently examined time slot (\refAlg{alg:fillschedulesegment}, lines 3 and 4), the \proc{Fill} algorithm will just schedule this video segment at maximum possible quality. This behavior ensures a minimum number of segments in the buffer as long as there is no need for buffering more segments for future time slots with insufficient data rate.
%If there is enough capacity in the current time slot , but
%the video playback has incurred a playback interruption before, the algorithm will try to download more segments to avoid further playback interruptions (\refAlg{alg:fillscheduler}, lines 23 to 29).

If, during the iteration, the anticipated data rate in some time slot $t$ does not suffice to download a new video segment (\refAlg{alg:fillschedulesegment}, lines 6 to 22), even at the lowest video quality level, the \proc{Fill} algorithm has to change the schedule for one
or more \emph{previous} time slots to download and buffer a video segment before time slot $t$ with insufficient data rate. This part of the algorithm, as outlined in \refAlg{alg:fillschedulesegment}, requires
the definition of the following helper functions: 
\begin{itemize*}
\item $\proc{getBestQuality}(Q,c)$\\
Returns the best downloadable quality (out of $Q$) for a segment with anticipated available data rate $c$, or FALSE if there is not enough data rate even for the lowest quality
\item $\proc{getBestQualityRange}(Q,n,c)$\\
Returns the best possible quality (out of $Q$) in which $n$ segments can be downloaded with anticipated available data rate $c$, or FALSE if there is not enough data rate to download even in the lowest quality
\item $\proc{getSegmentsForQuality}(q,c)$\\
Returns the number of downloadable segments with quality $q$ and available data rate $c$
% \item $\proc{getPossibleSegments}(u,t,l,C)$\\
% Returns how many out of $l$ segments feasible at a quality for user $u$ at time $t$ with capacities $C$
\item $\proc{schedule}(u,s,t,q)$\\
Schedule the download of segment $s$ for user $u$ at time $t$ with quality $q$
\end{itemize*}
These functions can be easily implemented and their precise implementation is omitted in this paper to improve the readability of the algorithm.

%The algorithm incrementally goes back through the previous time slots and checks if it is possible to download both the segments for the previous time slots and the initially checked time slot. If such a range of time slots is found, the \proc{Fill} algorithm reschedules all segments from that range to accomodate them and the new segment.

%The algorithm goes back time slot by time slot from time slot $t$ where a download of a full segment was not possible.
From time slot $t$ where a download of a full segment was not possible, the algorithm goes back time slot by time slot.
In these previous time slots, it downgrades the video quality of the segments, freeing up capacity to enable the download of the segment that has to be playout out in time slot $t$. It can push up the scheduled download times of earlier segments in order to fit more segments into time slots. Once a range of time slots is found where all segments including the one to be played out in time slot $t$ fit in (at reduced quality), the computation of the schedule up to time slot $t$ is complete. This schedule is then the basis to plan the download for the segment for time slot $t+1$ in the next iteration.

\begin{figure}[hbt]
  \begin{center}
  \includegraphics[width=0.315\textwidth]{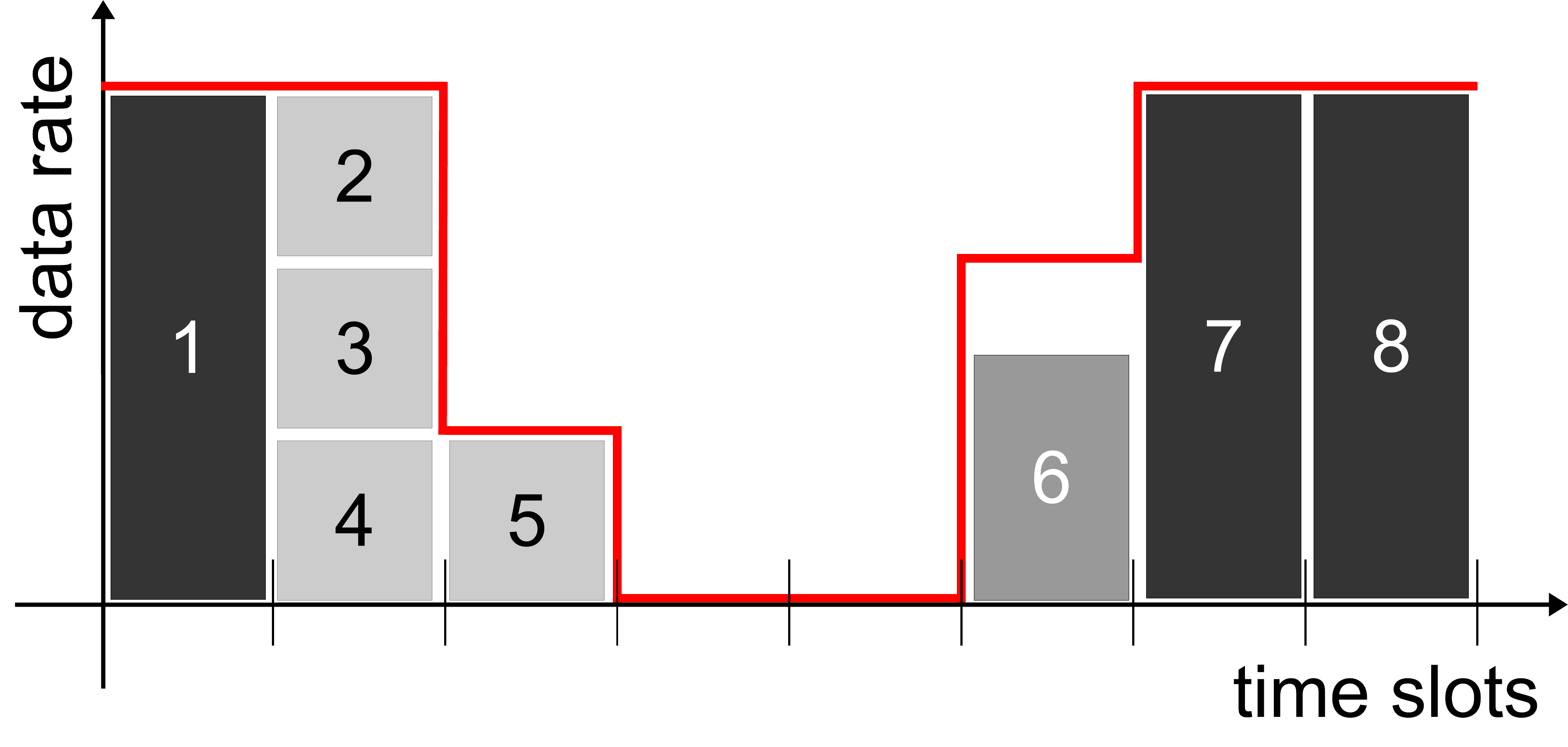}
  \caption{Example for Fill Algorithm}
  \label{fig:scheduler_fill}
  \end{center}
\end{figure}

For example, look at the example in \refFig{fig:scheduler_fill}, there is not enough data rate in the fourth and fifth time slots to download a video segment, but in the second time slot there is enough data rate to download three segments, so the algorithm will first go back
to the third time slot, determine that only going back to the third time slot is not sufficient and then also move back to the second time slot. By doing that the algorithm can resolve the lack of data rate in the fourth and fifth time slots and the video can be played back uninterruptedly.

With this behavior, the \proc{Fill} algorithm efficiently reduces the occurrence of playback interruptions. \refFig{fig:scheduler_fill} shows the resulting schedule from the \proc{Fill} algorithm
with the same example as for the previously described greedy scheduling algorithms: in the second time slot enough segments can be downloaded and buffered to play the video uninterruptedly. 

\begin{figure}[htb]
  \begin{center}
  \includegraphics[width=0.315\textwidth]{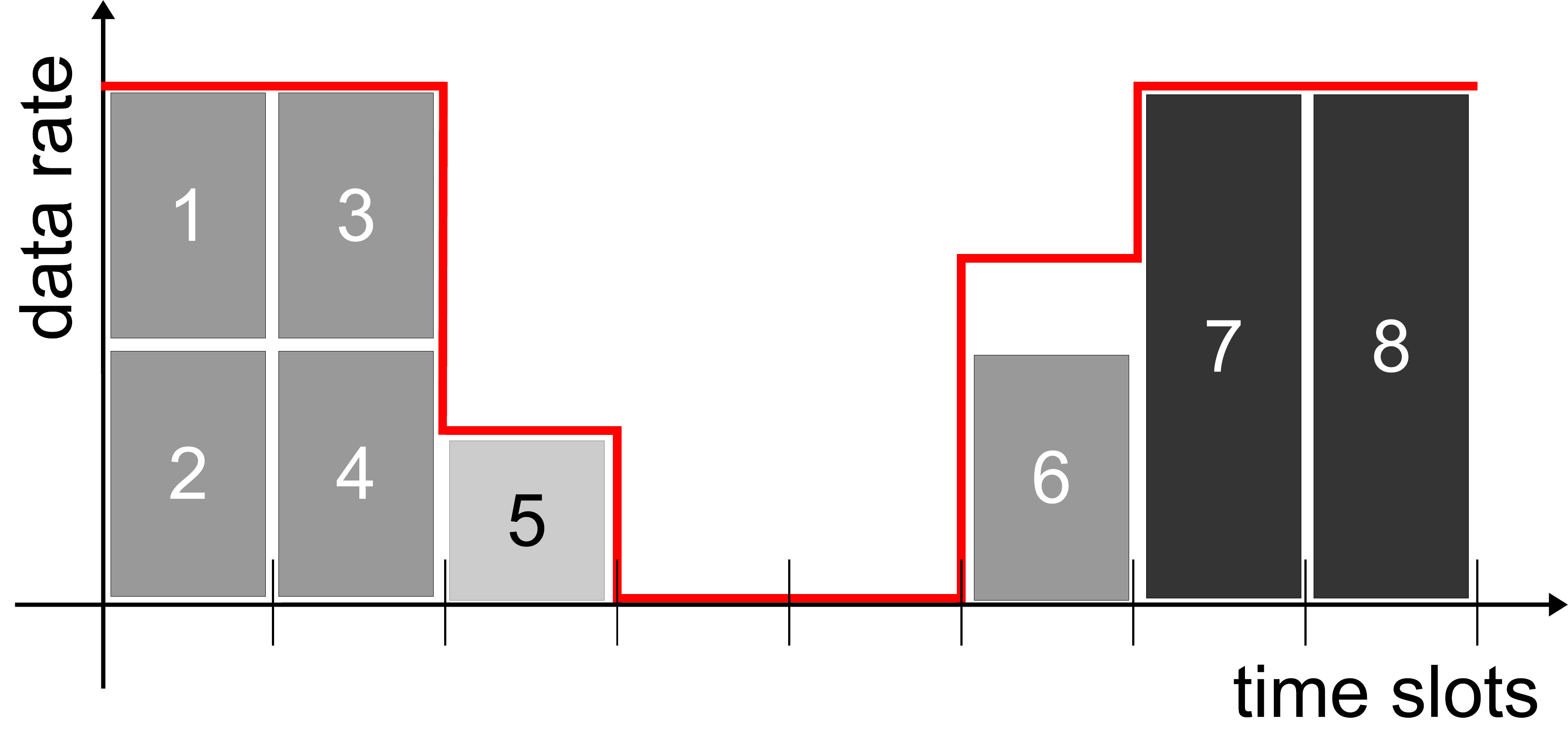}
  \caption{Example for MIQCP Schedule}
  \label{fig:scheduler_opt}
  \end{center}
\end{figure}

The downside of the \proc{Fill} algorithm is the fact that the reduction of playback interruptions also comes with a larger variance in the video quality level. By comparing the example schedule from the 
\proc{Fill} algorithm in \refFig{fig:scheduler_fill} with the schedule generated from the optimization problem in \refFig{fig:scheduler_opt} this behavior becomes obvious: the \proc{Fill} algorithm only
goes back to the second time slot and can resolve the lack of data rate in the fourth and fifth time slots by downloading the segments in the lowest video quality, whereas going back to the first time 
slot and downloading the segments with a medium video quality level would have provided the optimal average video quality level. In this toy example one could argue to allow the \proc{Fill} to go back 
a number of additional time slots to fix that issue. But in a real scenario there is no way to reasonably limit such a number of additional time slots to consider, thus the algorithm would not be
much different from a brute force algorithm with an excessive running time.

\section{System Design and Implementation}
\label{sec:system}
In this section we discuss how the previously introduced algorithms can
be integrated into a real system, using existing tools and extending existing 
protocols with backwards-compatible extensions as needed.
We first explain the design decisions and their implications on the system behavior and then continue with the system architecture and its interfaces in \refSec{sec:arch}.
We then explain in \refSec{sec:impl} implementation details and adjustments to the HLS protocol necessary to use the scheduling algorithms. The concrete Testbed implementation 
which we used to verify our simulation results is described afterwards in \refSec{sec:testbed}

\subsection{Design Decisions for Download Control}
The implementation of the previously introduced scheduling algorithms requires changes to an existing system
to control when which segments of a video are downloaded by user equipments (UEs). In order to implement these changes,
two design decisions with different advantages and disadvantages/costs have to be made:
\begin{itemize*}
  \item Should the buffering behavior be controlled at the UE or in the network?
  \item Should arbitrary or only preselected content providers be supported?
\end{itemize*}

% These design decisions have direct implications on the buffering behavior of the system:
% If the buffering is done too carefully, playback interruptions will occur when
% the preloaded buffer was too small (buffer under-runs).
% Even though too greedy buffering (over-buffering) does not degrade the user experience of the greedy UE, it could waste wireless resources which are needed by other UEs. Therefore we need to prevent greedy over-buffering.
These design decisions have direct implications on the buffering behavior of the system regarding over-buffering and buffer underruns.
For every combination of the design decisions we get the following requirements and capabilities of our system:

\begin{enumerate*}
\item \emph{No buffer control at UEs and arbitrary content providers}\\
This implementation requires deep packet inspection (DPI) on the network in
order to separate video traffic from other traffic, because we do not know
the content providers beforehand. That imposes additional cost and requires additional processing
for the network operator. When we do not have a modified UE which allows us to control the buffer,
we assume that the UE will be greedy. In order to implement our schedule we can only control the data flow in the network.
We can prevent over-buffering only by limiting the connection speed for a UE and therefore keep the UE from downloading more segments than it should.
But since we cannot force a UE to buffer more than it wants to, buffer underruns cannot be prevented.

\item \emph{No buffer control at UEs and preselected content providers}\\
When implementing the buffer control mechanisms for preselected content providers which we know beforehand only, the separation
of video traffic becomes trivial, in contrast to the previous case. The problem with buffer underruns however still remains the same.

\item \emph{Buffer control at UEs and arbitrary content providers}\\
When we have a means to control the buffer size of an UE, i.e.\ by a modified
version of the video player, we can prevent over-buffering and minimize buffer underruns by
explicitly instructing the UE from the scheduler how many segments it should load at a certain time.
A modified software could also support the separation of video traffic from other
traffic, i.e. by sending all video requests over a special proxy.

\item \emph{Buffer control at UEs and preselected content providers}\\
If we can fully control the buffering behavior and can easily separate video traffic from
other traffic the implementation of our system becomes most easy. We
then can optimize the buffer sizes on the UEs with little additional
complexity on the network side.
\end{enumerate*}

Although implementing our approach with the maximum level of control on both the UEs and the content providers is the easiest way, a
trade-off has to be taken here: Limiting the available content providers to a selected few also limits the usefulness to the users. However the best performance can only be achieved with modifications to the UE because otherwise we have no means to reliably prevent buffer underruns and preventing over-buffering is difficult to implement.

%\subsection{Architecture and Interfaces}
\subsection{Architecture and Implementation}
\label{sec:arch}
For our implementation we chose to use a modified video player on the UEs. With that we can fully control the buffer and we can analyze the performance of the system. Our implementation supports arbitrary content providers (in our tests we used our own video source to eliminate external influences).

\begin{figure}[htb]
%\begin{figure}[H]
  \begin{center}
  \includegraphics[width=0.45\textwidth]{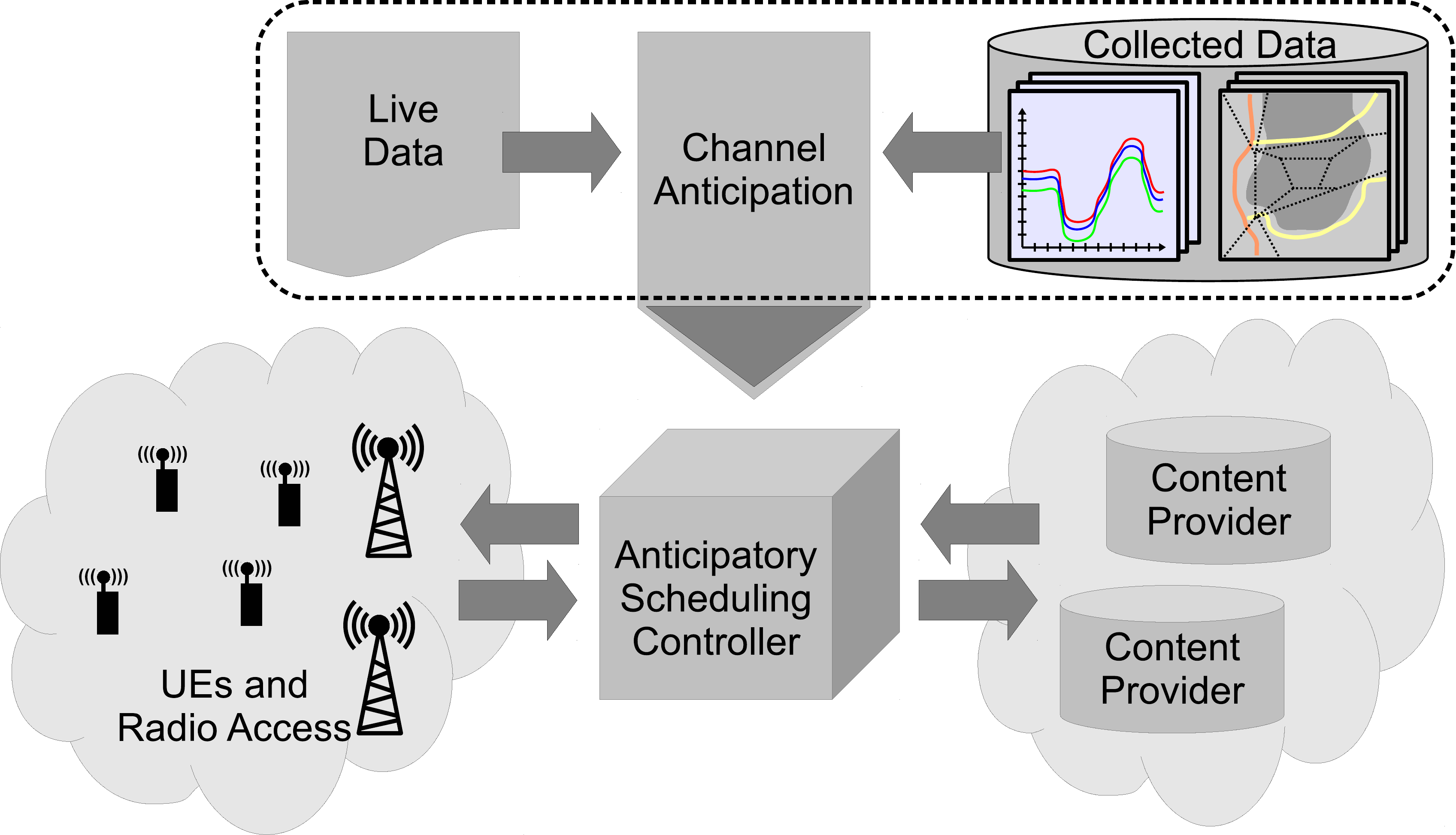}
  \caption{Architecture}
  \label{fig:architecture}
  \end{center}
\end{figure}

To implement our schedulers we assume an overall architecture as depicted in \refFig{fig:architecture}.
This architecture does not require any changes to current cellular radio interfaces and networks (RANs) and can be implemented in a cellular network as well as in a wireless LAN scenario, since the scheduler is implemented in higher layers.
It also does not require any changes to the content provider since all scheduling decisions
and the schedule is enforced in the \emph{Anticipatory Scheduling Controller}.

The \emph{Anticipatory Scheduling Controller}, as the central entity in this architecture, intercepts the requests from the UEs to the content providers. It can then perform the buffer control and quality selection with the following three steps:
\begin{enumerate*}
 \item Intercept the video request from the UE and analyze it (video data rates, available variants)
 \item Calculate schedule based on video data and anticipatory information on future data rates
 \item Control the buffering behavior of the UE according to the schedule
\end{enumerate*}

To do so, the \emph{Anticipatory Scheduling Controller} could be configured as an HTTP proxy (as HLS video requests are transported via HTTP). This could be enforced in cellular networks or done voluntarily by the users. Both operators and users have incentives to do so (less load on the network, better QoE for the users). 

The \emph{Channel Anticipation} works based on live data and previously collected data as explained in \refSec{sec:anticipation}. The anticipated data rates are then provided to the \emph{Anticipatory Scheduling Controller}.

\subsection{Protocol Extension}
\label{sec:impl}
We concentrated on HLS (HTTP Live Streaming)~\cite{hls} as the streaming protocol
for our implementation. It is available in the stock media players on Android \cite{hlsandroid} and iOS \cite{hlsapple} and is also available as an open-source implementation in the VLC player \cite{VLC}.
To stream a video using
HLS, regardless of using our extension or not, the video has to be encoded properly. This encoding is a CPU-intensive, one-time task. The video input 
is cut into independently playable segments with the same playback duration.
URLs to these segments are then added to a playlist. An example of such a normal HLS playlist is shown in \refFig{fig:hls_simple}.

\begin{figure}[htb]
\begin{minipage}[t]{0.225\textwidth}
\small
\begin{verbatim}
#EXTM3U
#EXT-X-VERSION:3
#EXT-X-TARGETDURATION:10
#EXTINF:10,
http://hostname/high/001.ts
#EXTINF:10,
http://hostname/high/002.ts
#EXTINF:10,
http://hostname/high/003.ts
#EXTINF:10,
http://hostname/high/004.ts
\end{verbatim}
\end{minipage}
\hspace{0.005\textwidth}
\vline
\hspace{0.005\textwidth}
\begin{minipage}[t]{0.225\textwidth}
\small
\begin{verbatim}
#EXTINF:10,
http://hostname/high/005.ts
#EXTINF:10,
http://hostname/high/006.ts
#EXTINF:10,
http://hostname/high/007.ts
#EXTINF:10,
http://hostname/high/008.ts
#EXT-X-ENDLIST
\end{verbatim}
\end{minipage}
\caption{Single variant HLS example with high quality segments, each 10 seconds long}
\label{fig:hls_simple}
\end{figure}

HLS streams can
provide multiple variants of the same video. Each variant can be encoded using
a different codec, bit-rate, or resolution.  HLS players can
switch between different variants because all segments have
equal length and are independently playable. A separate playlist is created for each variant
and additionally a master playlist with links to all variant playlists is used. An example of a master
playlist with three variants is shown in \refFig{fig:hls_master}. The master 
playlist contains parameters for each variant to enable HLS players to select the
most appropriate one. We use the \texttt{BANDWIDTH} parameter, given as a data rate in bit/s, for
each variant for this paper.
The created playlists and segments can then be placed on an HTTP server. An HLS player only needs the URL to the HLS master playlist. From there, all
variants and their segments are accessible.

\begin{figure}[tb]
\small
\begin{verbatim}
#EXTM3U
#EXT-X-STREAM-INF:BANDWIDTH=1000000
http://hostname/low/hls.m3u8
#EXT-X-STREAM-INF:BANDWIDTH=1500000
http://hostname/med/hls.m3u8
#EXT-X-STREAM-INF:BANDWIDTH=3000000
http://hostname/high/hls.m3u8
\end{verbatim}
\caption{Multi-variant master playlist with three variants}
\label{fig:hls_master}
\end{figure}

To control the buffering behavior of HLS players, we need a method to pass messages
to them. HLS players have no interface to receive control data besides playlists and segments via their own HTTP-GET requests.
We intercept the requests for playlists and modify the replies in the anticipatory scheduling controller.

The controller is aware of the schedule but also needs a means of 
inserting buffering instructions in the playlists. Thus, we introduce two 
new tags to HLS playlists: \texttt{BUFFERSIZE} and \texttt{REFRESH}. 
Both are defined as natural numbers including 0. These new tags are 
backwards compatible because the HLS standard instructs players to ignore tags which they 
do not recognize \cite{hls}.

\texttt{BUFFERSIZE} sets the size of the HLS player buffer
to the given value. Up to this amount of segments, the player will just greedily try to download more segments. If there are more segments in the buffer than instructed,
the buffer content is played, and no downloaded segments are discarded.
As soon as there are fewer segments in the buffer than the given limit, the
HLS player downloads additional segments to fill the buffer.

The \texttt{REFRESH} parameter instructs the HLS player to refresh the
playlists every \texttt{REFRESH} seconds. This will then update the 
\texttt{BUFFERSIZE} and \texttt{REFRESH} parameters. 
We suggest to set \texttt{REFRESH} to the playback length of a segment, thus after playing one segment the
HLS player will update its buffering parameters.

The two parameters together
solve the over-buffering and buffer underrun problem by precisely adapting the HLS player buffer size according to the schedule. This indirectly
influences \emph{when} an HLS player can download a segment.

Another property of an HLS stream that the scheduling algorithm needs to decide is
\emph{which quality} to download. In the case of multi-variant HLS streams, the player would
try to download the segments in the quality it prefers by doing its own local measurements.
But the schedules also include the HLS video quality for each segment, selected from the available
HLS variants.

Every time the HLS player requests an HLS master playlist the anticipatory scheduling controller downloads
the playlists of the scheduled variants and creates a single variant playlist out
of the multi-variant playlist. As shown in \refFig{fig:hls_example_joined}, segments from different 
variants are being selected and placed in a new single variant playlist according to
the example schedule in \refFig{fig:scheduler_opt}.
Only the joined (single-variant) playlist is then returned to the HLS player.
The decision which quality to download is hereby made by the anticipatory scheduling controller and not by the player anymore.
The joined playlist contains the \texttt{REFRESH} and \texttt{BUFFERSIZE} parameters. 
Each time a player refreshes an HLS playlist, it can receive a different value
for the \texttt{BUFFERSIZE} parameter. 
The values for each time slot of the \texttt{BUFFERSIZE} in \refFig{fig:hls_example_joined} are listed in \refTab{tab:buffer_size_values} for each refresh of the playlist.

\begin{figure}[htb]
\begin{minipage}[t]{0.225\textwidth}
\small
\begin{verbatim}
#EXTM3U
#EXT-X-VERSION:3
#EXT-X-TARGETDURATION:10
#EXT-X-BUFFERSIZE: 2
#EXT-X-REFRESH:10
#EXTINF:10,
http://hostname/med/001.ts
#EXTINF:10,
http://hostname/med/002.ts
#EXTINF:10,
http://hostname/med/003.ts
\end{verbatim}
\end{minipage}
\hspace{0.005\textwidth}
\vline
\hspace{0.005\textwidth}
\begin{minipage}[t]{0.225\textwidth}
\small
\begin{verbatim}
#EXTINF:10,
http://hostname/med/004.ts
#EXTINF:10,
http://hostname/low/005.ts
#EXTINF:10,
http://hostname/med/006.ts
#EXTINF:10,
http://hostname/high/007.ts
#EXTINF:10,
http://hostname/high/008.ts
#EXT-X-ENDLIST
\end{verbatim}
\end{minipage}
\caption{Joined playlist using the MIQCP Schedule from \refFig{fig:scheduler_opt} with \texttt{REFRESH} and \texttt{BUFFERSIZE} extensions (\texttt{BUFFERSIZE} set for time slot 1)}
\label{fig:hls_example_joined}
\end{figure}

\begin{table}[htb]
\tih{MD: \$TBV heisst jetzt $\beta$, und das muss auch so, da das \texttt{BUFFERSIZE} pro Zeit ist}
\caption{\texttt{BUFFERSIZE} values for time slots}
\label{tab:buffer_size_values}
\centering
\begin{tabular}{rcccccccc}
\toprule
Time slot & 1 & 2 & 3 & 4 & 5 & 6 & 7 & 8 \\\midrule
\texttt{BUFFERSIZE} & 2 & 3 & 3 & 0 & 0 & 1 & 1 & 1\\
\bottomrule
\end{tabular}
\end{table}

Through both mechanisms, the buffer size (\emph{when} to download) and preselection of variants (\emph{which quality} to download)
%, the when in time do play which HLS variant
can be controlled.
Thus, anticipatory 
buffering and variant selection based on the previously described algorithms
can be performed by simply extending the HLS protocol with two small extensions to the playlist parameters. 

\subsection{Testbed}
\label{sec:testbed}
% Simulating different algorithms to compare their performance is a good way to try out different things quickly
% without the need to have a real system at hand. But simulations can only show part of the picture
% since they are based on models of the real world and therefore rely on simplifying assumptions,
% which are not necessary true in a real system.
In order to analyze our algorithms and to test our HLS protocol extension in a real system,
we developed a testbed that allows us to run extensive tests with real hardware and compare the results of 
these tests with our simulations. We describe our testbed setup here and will present the simulation and testbed measurement results in \refSec{sec:eval}.

The testbed is based on the general architecture explained before. 
The UEs are smartphones and tablets with a customized Android operating system and a modified VLC video player.
Our modifications enable VLC to parse the additional playlist parameters and adapt its
buffer size accordingly. It also outputs extended information about the
buffer size and the downloaded segments which is used for our measurements. 

% Since the deployment of an long-term evolution environment (LTE) was neither feasible nor necessary for our tests,
% the radio access is based on plain wireless LAN, provided by four access points.
%These access points are normal PCs with wireless LAN cards and Linux with hostapd running on them.
The radio access in the testbed is implemented with 802.11g wireless LAN \cite{wlan} without any modifications and four access points. As explained before, the scheduling only happens
on the application layer, thus changes to the wireless MAC are not necessary. The access points are normal PCs with wireless LAN cards and Linux with hostapd running on them.

\begin{figure}[htb]
  \begin{center}
  \subfloat[Testbed Architecture]{
  \includegraphics[width=0.45\textwidth]{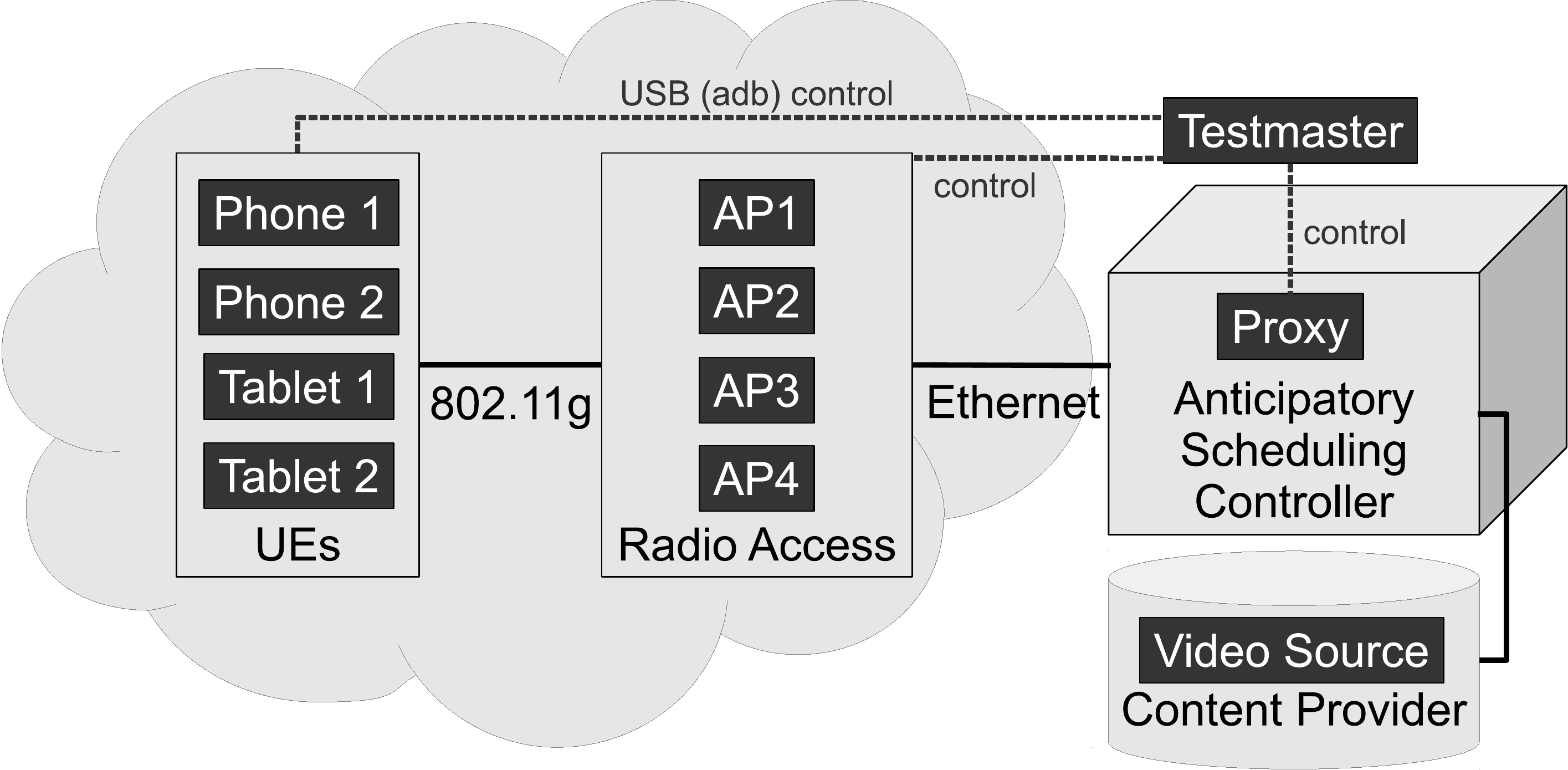}
  \label{fig:tb_architecture}
  }
  \\
  \subfloat[Testbed Setup]{
  \includegraphics[width=0.4750\textwidth]{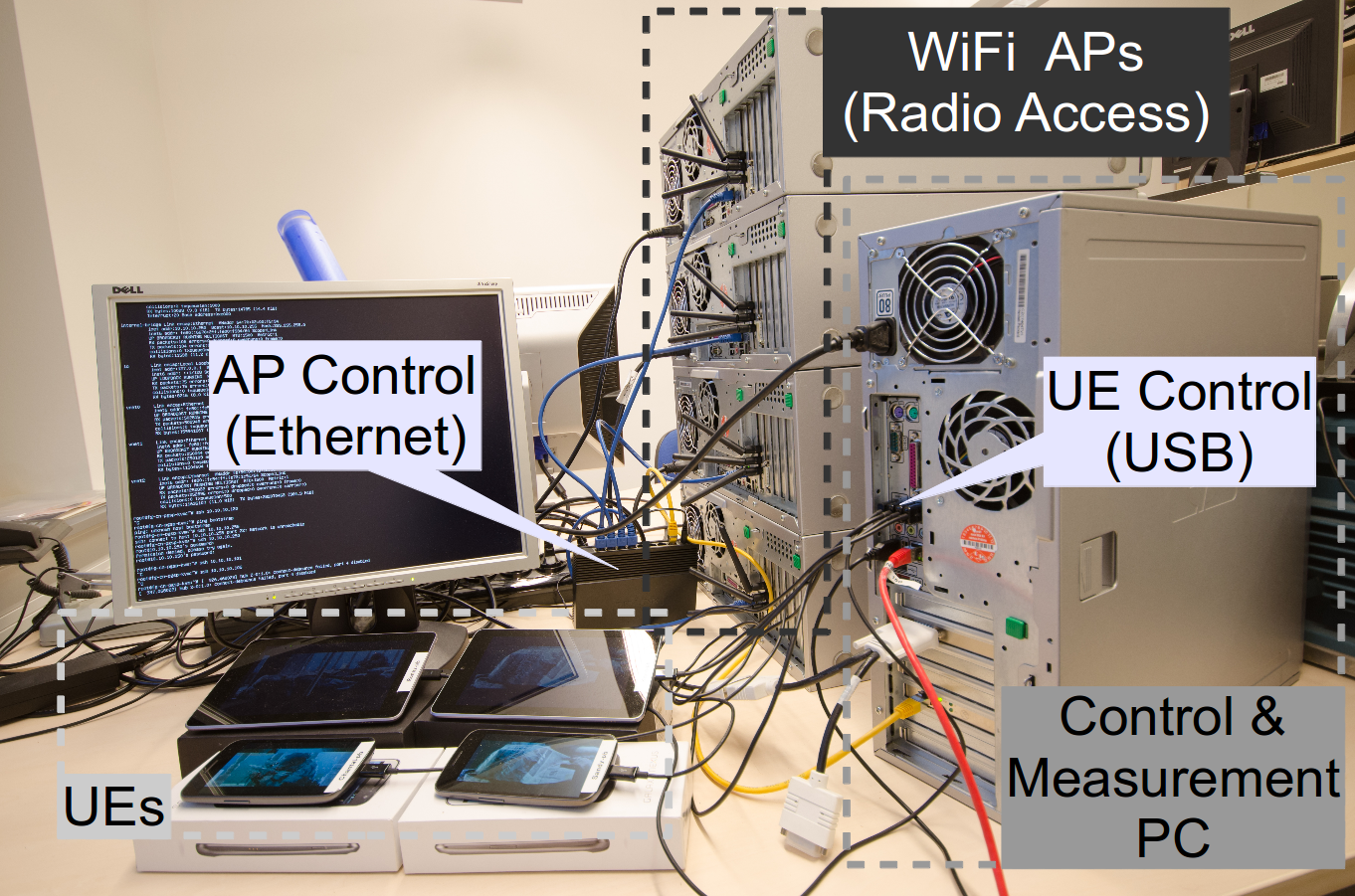}
  \label{fig:tb_photo}
  }
  \caption{Testbed}
  \label{fig:tb_all}
  \end{center}
\end{figure}

A fifth PC serves as central control and measurement unit and as a host for running the
anticipatory scheduling controller. All phones are connected to this PC via USB and are controlled with the 
Android debug bridge (ADB); the access points are controlled via an SSH connection. With the ADB we execute arbitrary shell commands on the phones
and emulate simple user interaction like starting or stopping a video stream. No data is transmitted via USB; it only serves to make experiments repeatable.
%The resulting testbed can be seen in \refFig{fig:tb_photo}.
%The resulting overall testbed architecture can be seen in \refFig{fig:tb_architecture} and a photo of the testbed in \refFig{fig:tb_photo}.
The resulting overall testbed architecture and setup can be seen in \refFig{fig:tb_all}.

% \begin{figure}[bht]
% %\begin{figure}[H]
%   \begin{center}
%   \includegraphics[width=0.45\textwidth]{figures/testbed_architecture2.pdf}
%   \caption{Testbed Architecture}
%   \label{fig:tb_architecture}
%   \end{center}
% \end{figure}
% \begin{figure}[bht]
% %\begin{figure}[H]
%   \begin{center}
%   \includegraphics[width=0.4750\textwidth]{figures/testbed.png}
%   \caption{Testbed}
%   \label{fig:tb_photo}
%   \end{center}
% \end{figure}

For the  HLS video stream content we used the publicly available movie ``Tears Of Steal''\footnote{\url{http://www.tearsofsteel.org/}} which we converted
using the VLC framework. The segments and playlists are served by an unmodified Apache webserver.

The anticipatory scheduling controller, which intercepts and modifies the
playlist requests from the UEs, is implemented as a transparent HTTP proxy
using the Python framework Twisted \cite{twisted}. The access points redirect all traffic
coming from the UEs to the proxy thus it is not necessary to change any
preferences on the UEs.

We wanted to be able to run a lot of repeatable and comparable tests, which
is why the movement of the UEs is emulated and not done physically. Movement emulation works by limiting the link speed and enforcing handovers between access points.
We achieve this by using standard traffic shaping capabilities of Linux
on the access points and on the phone. From a predefined scenario we get the
data rate for every UE and base station per time slot. These values are
then set as speed limits on our access points at the corresponding time. 
Handover events between the access points are also precalculated from the scenario and then triggered on the
phones. With this setup we can run tests without the need to physically move the UEs.

We automatically start the video stream via the ADB connection to the phones and collect information about the streaming
(i.e.\ \emph{when} a segment has been actually loaded in \emph{which
quality}). The results returned by the testbed runs are in the same format as the
simulation results and allow a direct comparison.

%\section{Simulation and Measurement Results}
\section{Simulation and Testbed Results}
\label{sec:eval}
In this section we present both simulation results and results from measurements with the previously described testbed. For the simulation we use our own Python implementation.
Before presenting the results we define the evaluation scenarios.

\subsection{Scenario}
\begin{figure}[htb]
  \begin{center}
    \includegraphics[width=0.475\textwidth]{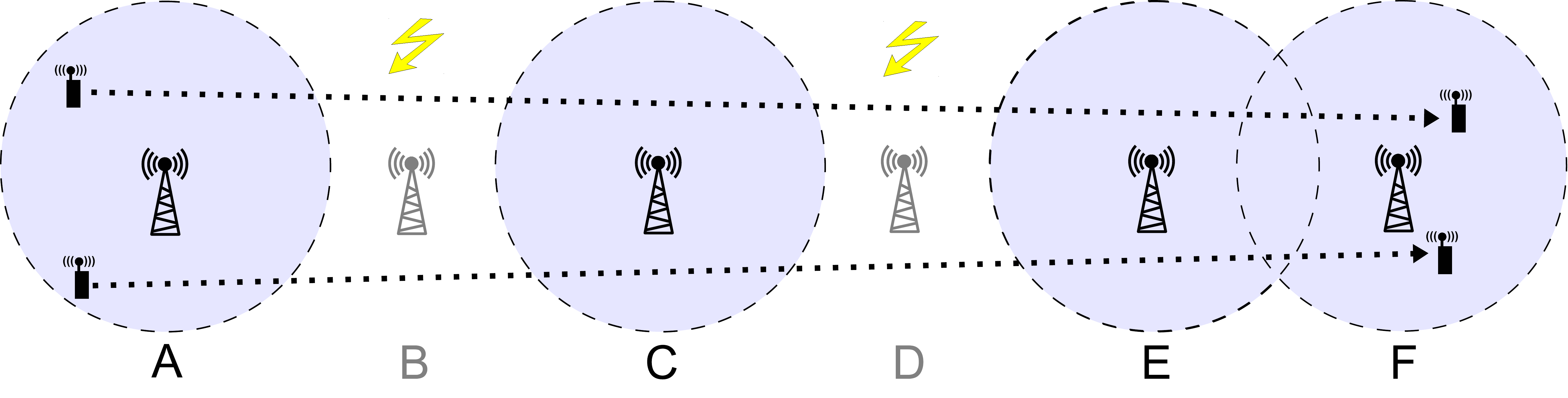}
    \caption{Scenario}
    \label{fig:sim_scenario}
  \end{center}
\end{figure}

The basic structure for the evaluation scenario, for both simulation and testbed measurements, is a line of base stations with the users moving through the scenario from the
first base station to the last base station as illustrated in \refFig{fig:sim_scenario}.
To reduce the available data rate and to create the need for buffering, we remove cells from the scenario, as
illustrated with base stations B and D. The more cells we remove, the more gaps without any available data rate occur and the more segments have to be buffered to avoid playback interruptions. The users all move as a group
from the first base station to the last base station (e.g., a train scenario).
Apart from the pattern in which the base stations are removed, the scenario parameters for the simulation and the testbed measurements are the
same.%, thus both results can be compared.

The wireless radio is modeled according to 3GPP Long Term Evolution (LTE) \cite{ltetr}. The base stations are placed equidistantly with an inter-site distance of 1500 meters,
which is slightly larger than a normal urban scenario in order to augment the effects resulting from removing cells. We consider four active users in the scenario because the testbed
setup only contains four devices and we want to maintain comparability between the simulation and the testbed measurements.

The path loss in dB between the base stations and the users is obtained by $128.1 + 37.6 \cdot \log_{10}(d) + S_{ln}$ \cite{ltetr}, where $d$ represents the distance between the base station and
the user in kilometers and $S_\mathrm{ln}$ is a normal random variable with zero mean and standard deviation of 10\,dB to model slow fading.

For the channel capacity we assume an asymptotically error-free communication channel, modeled by the Shannon equation with the parameters listed in \refTab{tab:shannon}.
The maximum data rate for a base station is limited to 30\,Mbit/s to account for the small number of users in the scenario. The allocation of data rates to the users
in each time slot is up to a wireless resource scheduler, which is in our case a simple proportional fair scheduler.
%For each time slot each user is associated with the base station that provides the best channel capacity.

\begin{table}[htb]
\centering
\caption{Evaluation Parameters}\label{tab:shannon}
\begin{tabular}{lr}\toprule
Channel bandwidth & 10\,MHz\\
Transmit power & 46\,dBm\\
% Antenna gain & 0\,dB\\
% Antenna & isotropic\\
Antenna & isotropic, 0\,dB gain\\
Noise PSD & -174\,dBm/Hz\\
%Noise floor & 9 dBm\
%Average Interference & -149.0138 dBm/Hz\\
Average Interference & -149\,dBm/Hz\\
\midrule
Inter site distance & 1500\,m\\
Number of users & 4\\
Number of base stations & 44\\\bottomrule
%Capacity per base station & 100\,Mbit/s\\
%Segment length & 10\,s\\
%Number of segments & 10\\\bottomrule
\end{tabular}
\end{table}

\begin{figure*}[tb]
%\begin{figure}[H]
  \begin{center}
  \subfloat[Average Quality]{
  \includegraphics[width=0.28\textwidth]{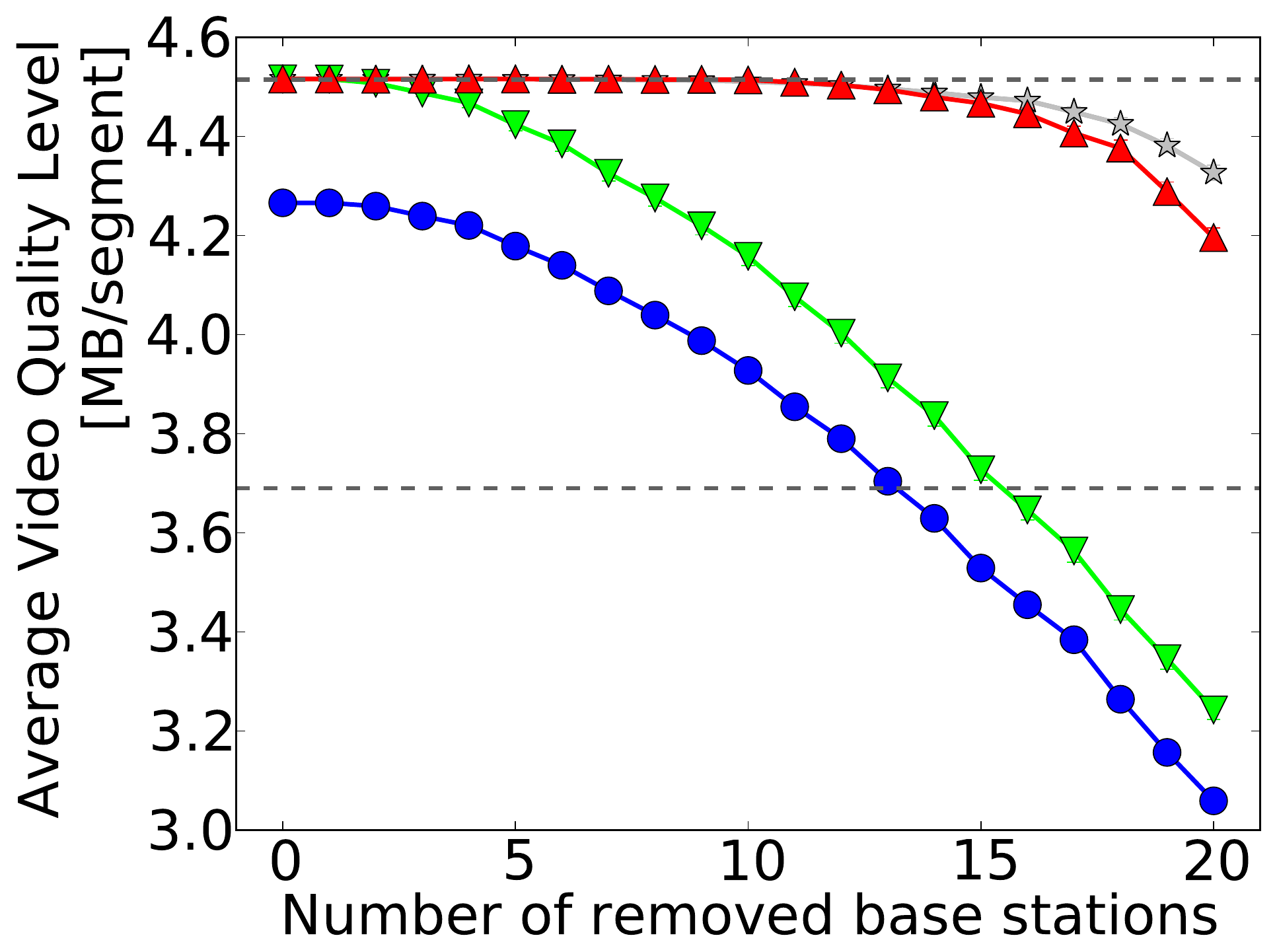}
  \label{fig:sim_quality}
  }
  %\\
  \subfloat[Average Lateness]{
  \includegraphics[width=0.28\textwidth]{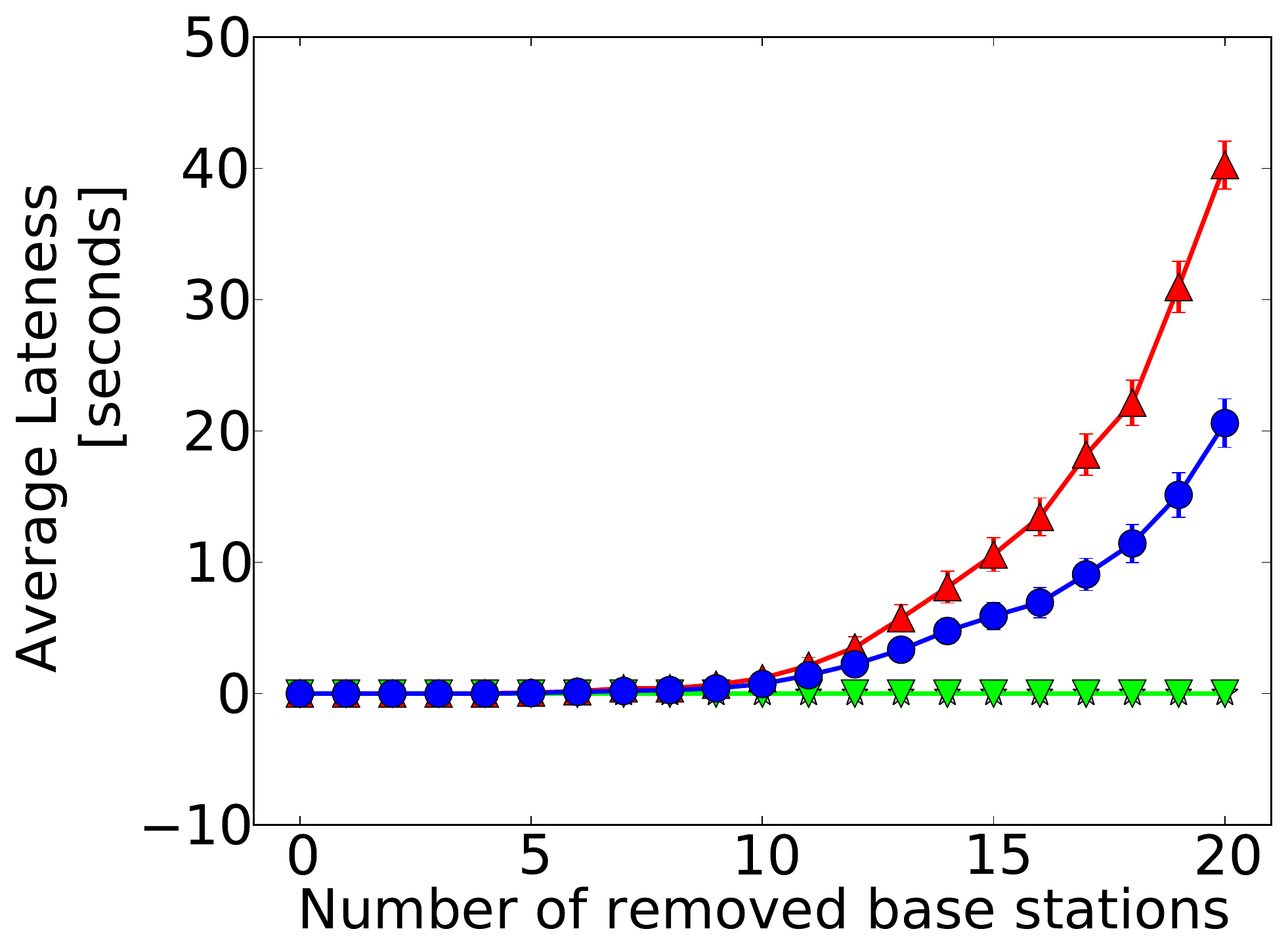}
  \label{fig:sim_lateness}
  }
    %\\
   \subfloat[Average Buffering]{
   \includegraphics[width=0.28\textwidth]{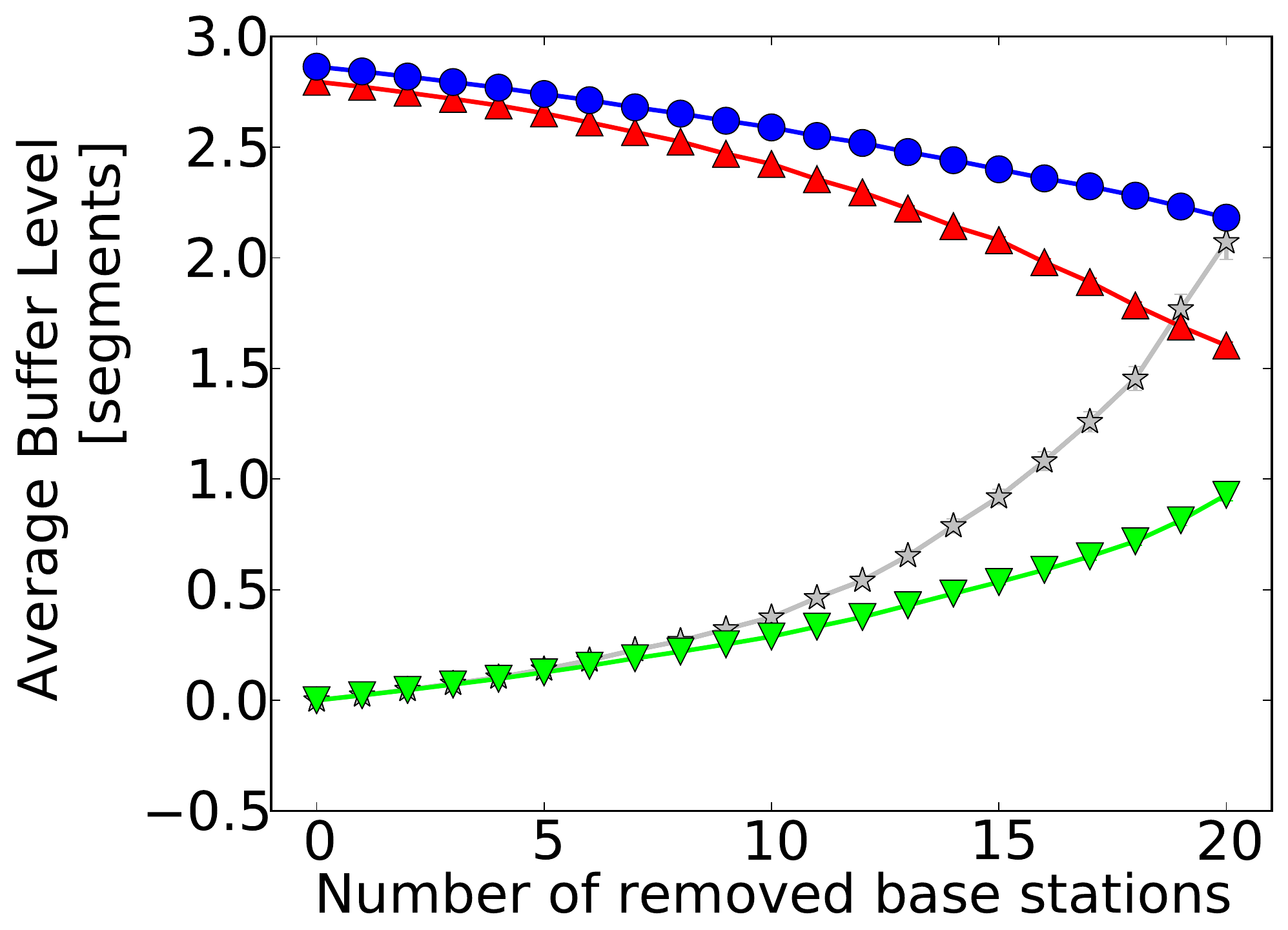}
   \label{fig:sim_buffer}
   }%\\
  \subfloat{
  \includegraphics[width=0.1\textwidth]{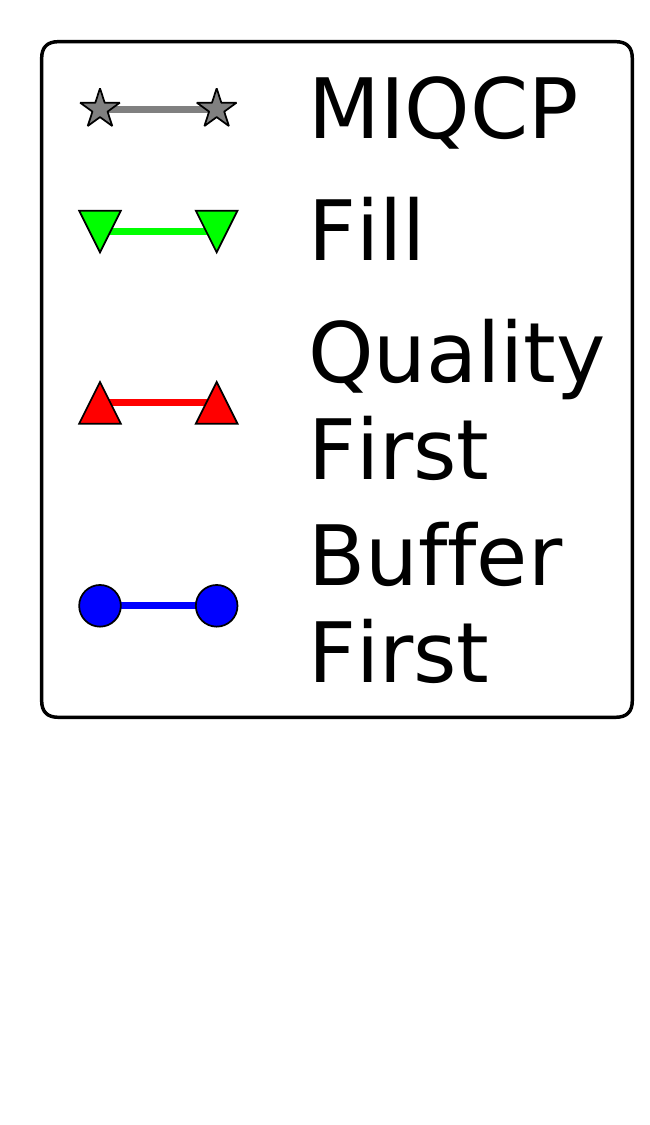}
  \label{fig:sim_legend}
  }
  \caption{Simulation Results}
  \label{fig:sim}
  \end{center}
\end{figure*}

%\ti{MD: weights explzit angeben?}
The weights for the MIQCP objective function, as described in \refSec{sec:optimization}, are set to enforce the following lexicographical order:
minimize lateness before maximizing quality and before minimizing buffering ($W_l=440$, $W_q=10$, $W_b=1$). The maximum buffer size for
the greedy scheduling algorithms is set to $3$ segments, which corresponds to the default setting for VLC on Android. 

The video quality levels and the resulting required data rates are taken from the test video we generated from the clip ``Tears of Steel''.
%using the VideoLAN player as the encoder.
%The resulting segment sizes are listed in \refTab{tab:vquali}.
The resulting segment sizes for the three video quality levels are 1.77\,MB (low), 3.69\,MB (medium) and 4.51\,MB (high).
%The scheduling algorithms use one segment size for each video quality level, whereas the real file size of all segments varies slightly by a few hundred kilobytes due to the video encoding. Thus we used the maximum segment size over all generated video segments in one video quality level as the parameter for the scheduling algorithms.
As the real file size of all segments varies slightly by a few hundred kilobytes due to the video encoding, we use the maximum size over all generated video segments in one video quality level as the parameter for the scheduling algorithms.
%We use a segment length of 10 seconds, which corresponds to the recommended segment length in the HLS standard.
We use a segment length of 10 seconds, corresponding to the recommended value in the HLS standard.

% \begin{table}[H]
% \centering
% \caption{Video Quality Levels}\label{tab:vquali}~\\
% \begin{tabular}{lrr}\toprule
% quality & resolution & MB per\\
% level & & segment\\\midrule
% Low & 1280x720 & 1.77 \\
% Medium & 1280x720 & 3.69\\
% High & 1280x720 & 4.51\\\bottomrule
% \end{tabular}
% \end{table}

\subsubsection{Simulation Scenario}

For the simulation scenario we are not limited to the number of physical devices we have in the testbed. Thus we use a total of $44$ base stations and a video length of $44$ segments. 
%\ti{MD: warum gleich?}

To induce the need for buffering we randomly remove base stations from the scenario. The number of removed base stations varies from $0$ to $20$, which means that in the worst case half of all
base stations are removed. The removed base stations are selected uniformly, whereas the first and last $2$ of the $44$ base stations are never removed to avoid side effects.
Removing more base stations yields infeasible scenarios for MIQCP, because some segments can never be downloaded and thus violate the constraints.

\subsubsection{Testbed Scenario}

In the testbed, which we described in \refSec{sec:testbed}, the scenario is limited by the number of physical devices in the testbed. We have again $4$ users, the phones and tablets in the testbed,
but in contrast to the simulation only $4$ base stations. The base stations are again arranged in a line but with only one fixed gap without any available data rate in the middle
.%, as shown in \refFig{fig:testbed_scenario}.
In order to vary the need for buffering we perform measurements with a gap equal to the range of $2$ and $4$ base stations. 

% \begin{figure}[htb]
%   \begin{center}
%     \includegraphics[width=0.45\textwidth]{figures/testbed_scenario.pdf}
%     \caption{Testbed Scenario}
%     \label{fig:testbed_scenario}
%   \end{center}
% \end{figure}

\subsection{Results}

\begin{figure*}[thb]
%\begin{figure}[H]
  \begin{center}
    \subfloat{
    \includegraphics[width=0.28\textwidth]{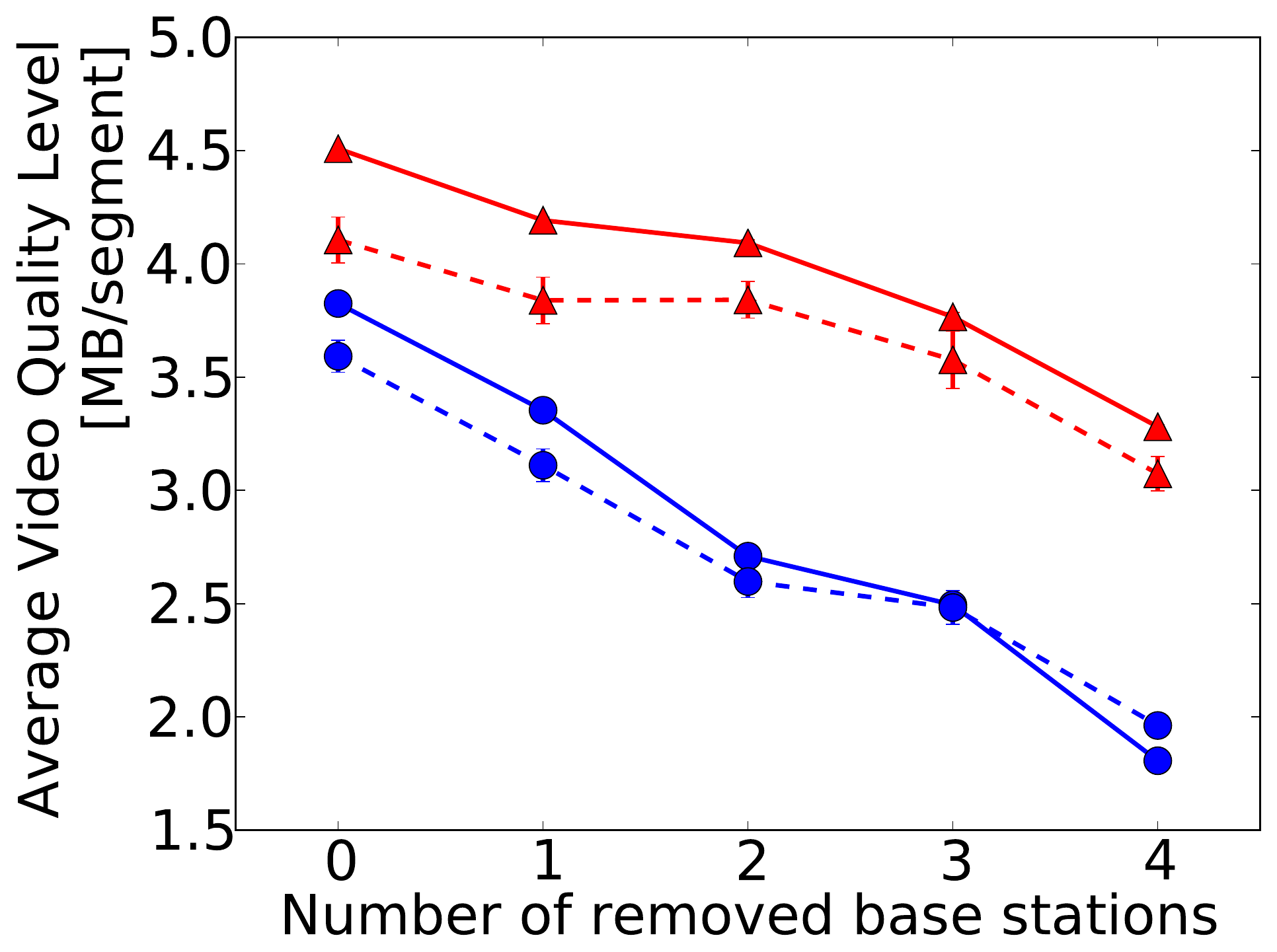}
  \label{fig:tb_qualityx}
  }
  %\\
  \subfloat{
  \includegraphics[width=0.28\textwidth]{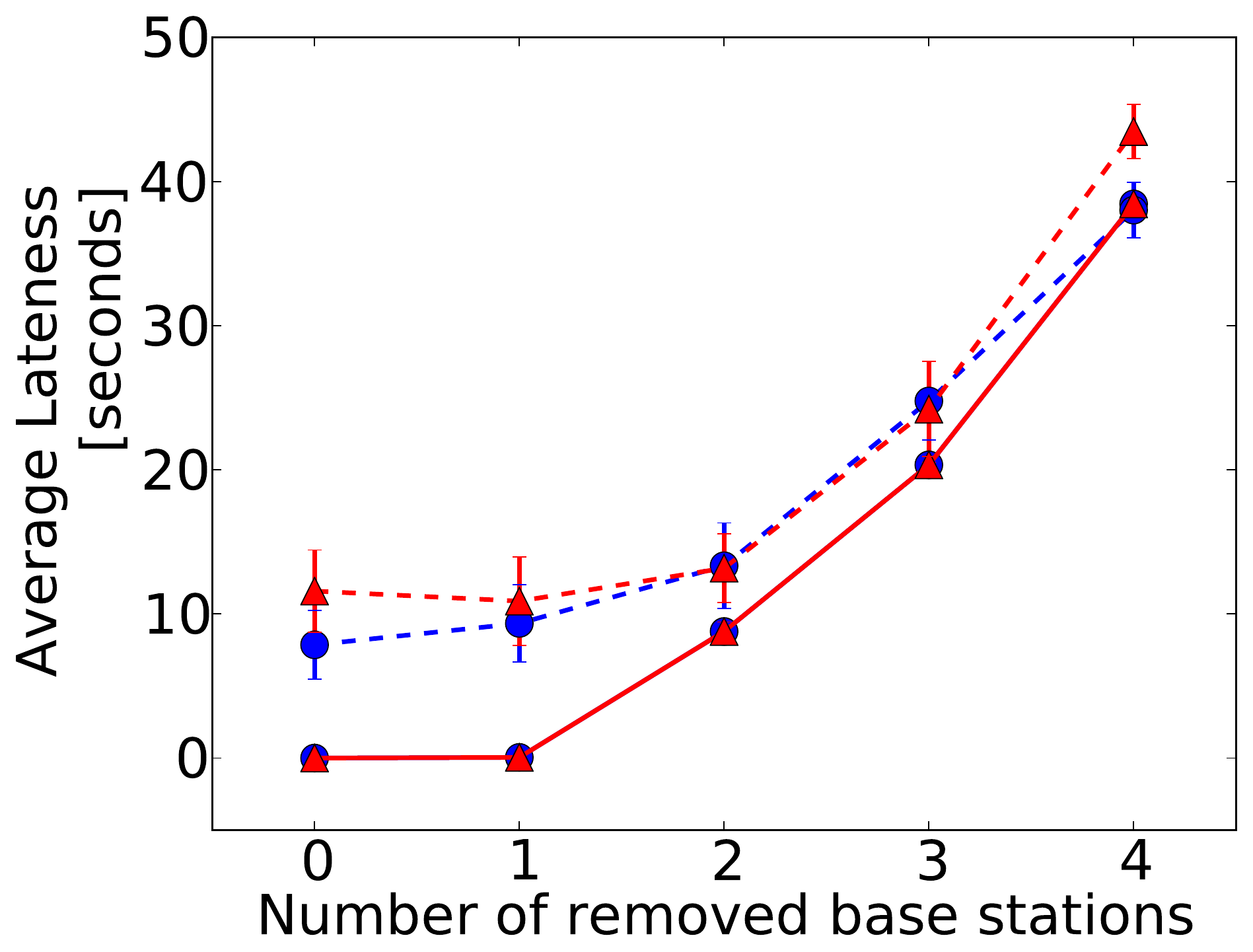}
  \label{fig:tb_latenessx}
  }
    %\\
   \subfloat{
   \includegraphics[width=0.28\textwidth]{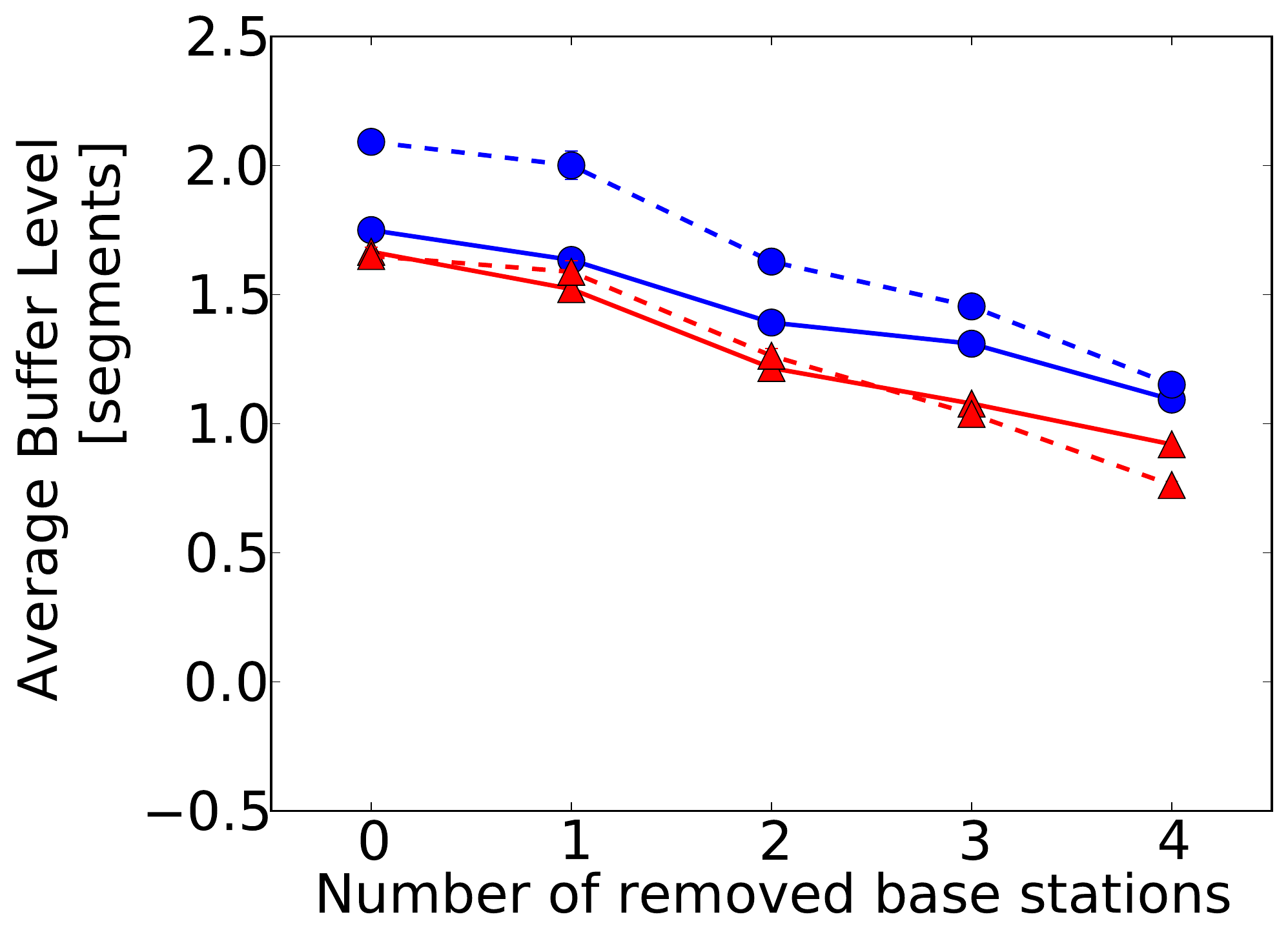}
   \label{fig:tb_bufferx}
   }
  \subfloat{
  \includegraphics[width=0.12\textwidth]{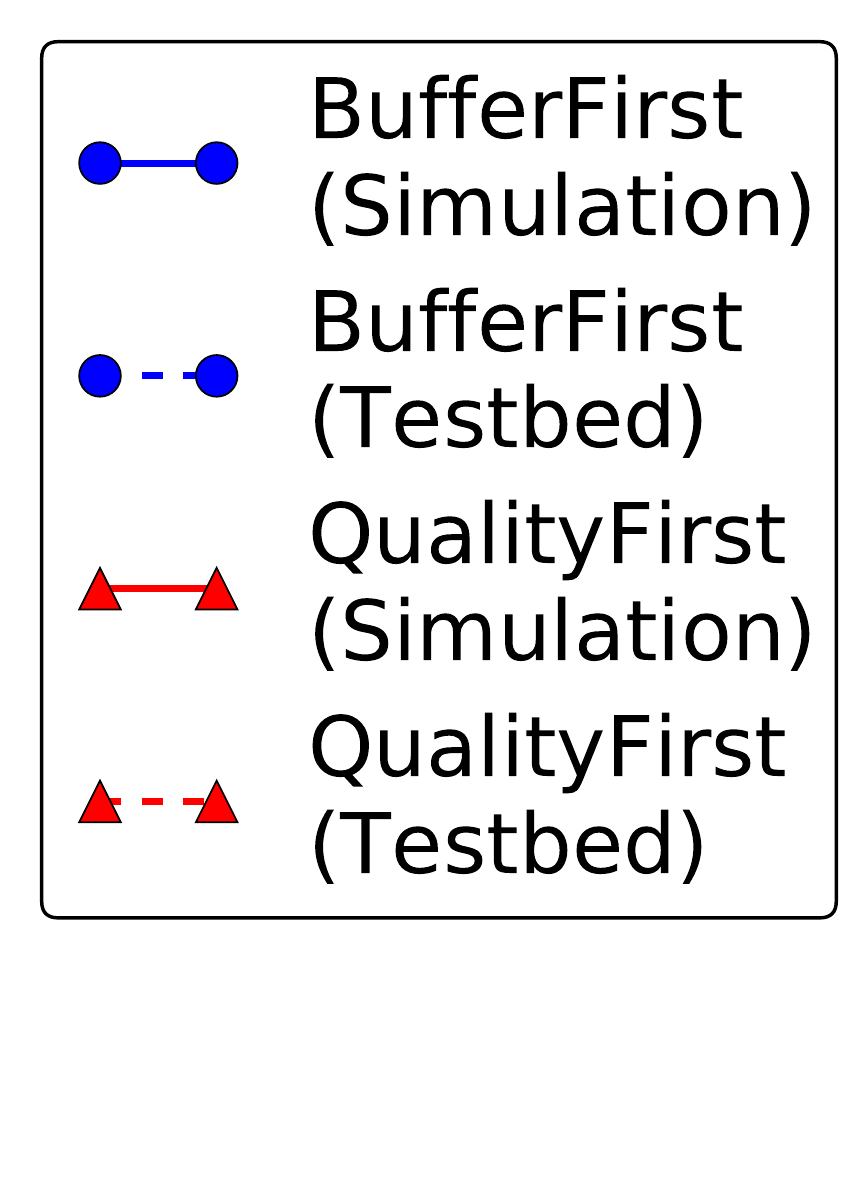}
  \label{fig:tb_legendx}
  }\\
  \setcounter{subfigure}{0}
  \subfloat[Average Quality]{
  \includegraphics[width=0.28\textwidth]{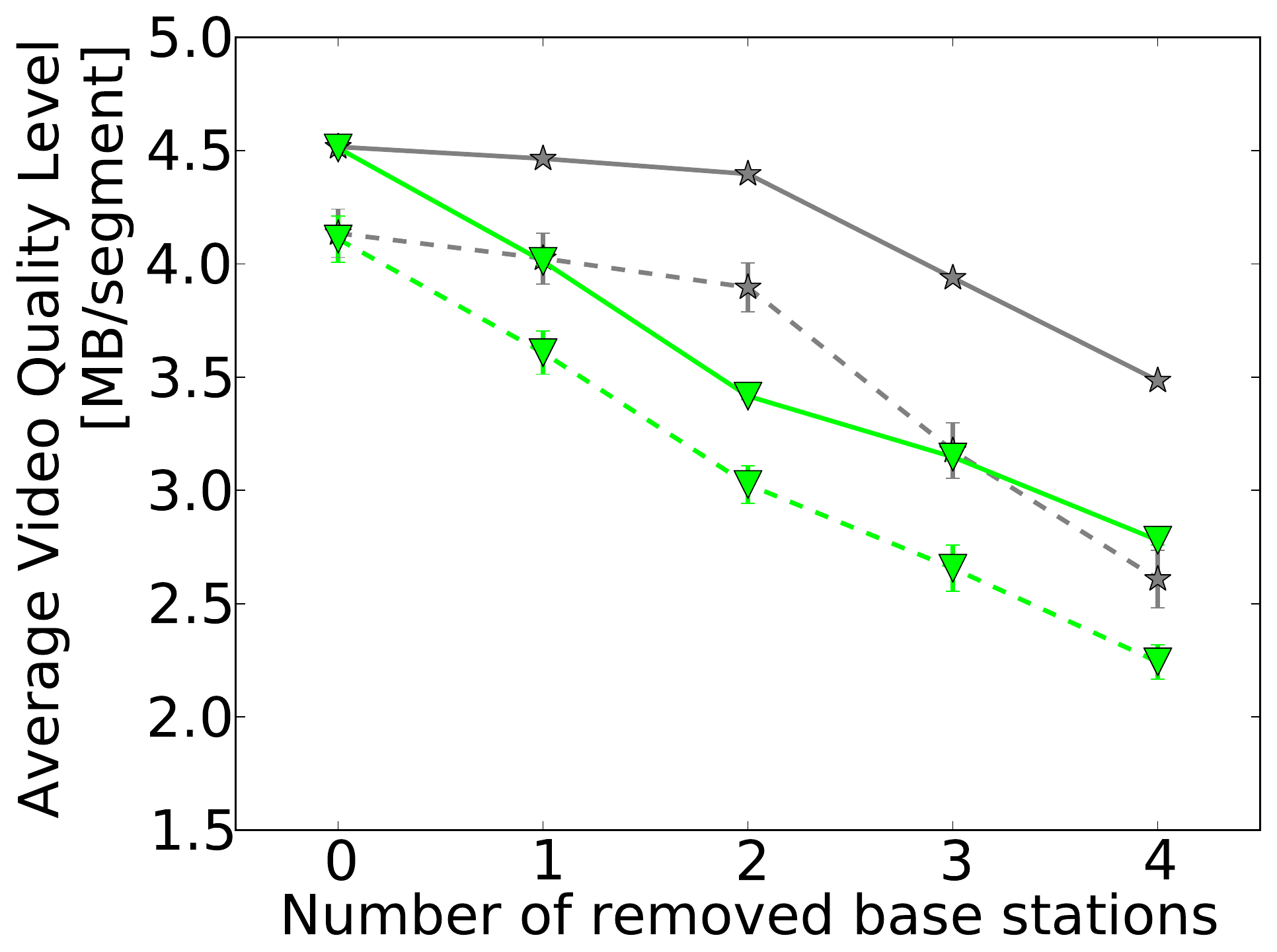}
  \label{fig:tb_quality}
  }
  %\\
  \subfloat[Average Lateness]{
  \includegraphics[width=0.28\textwidth]{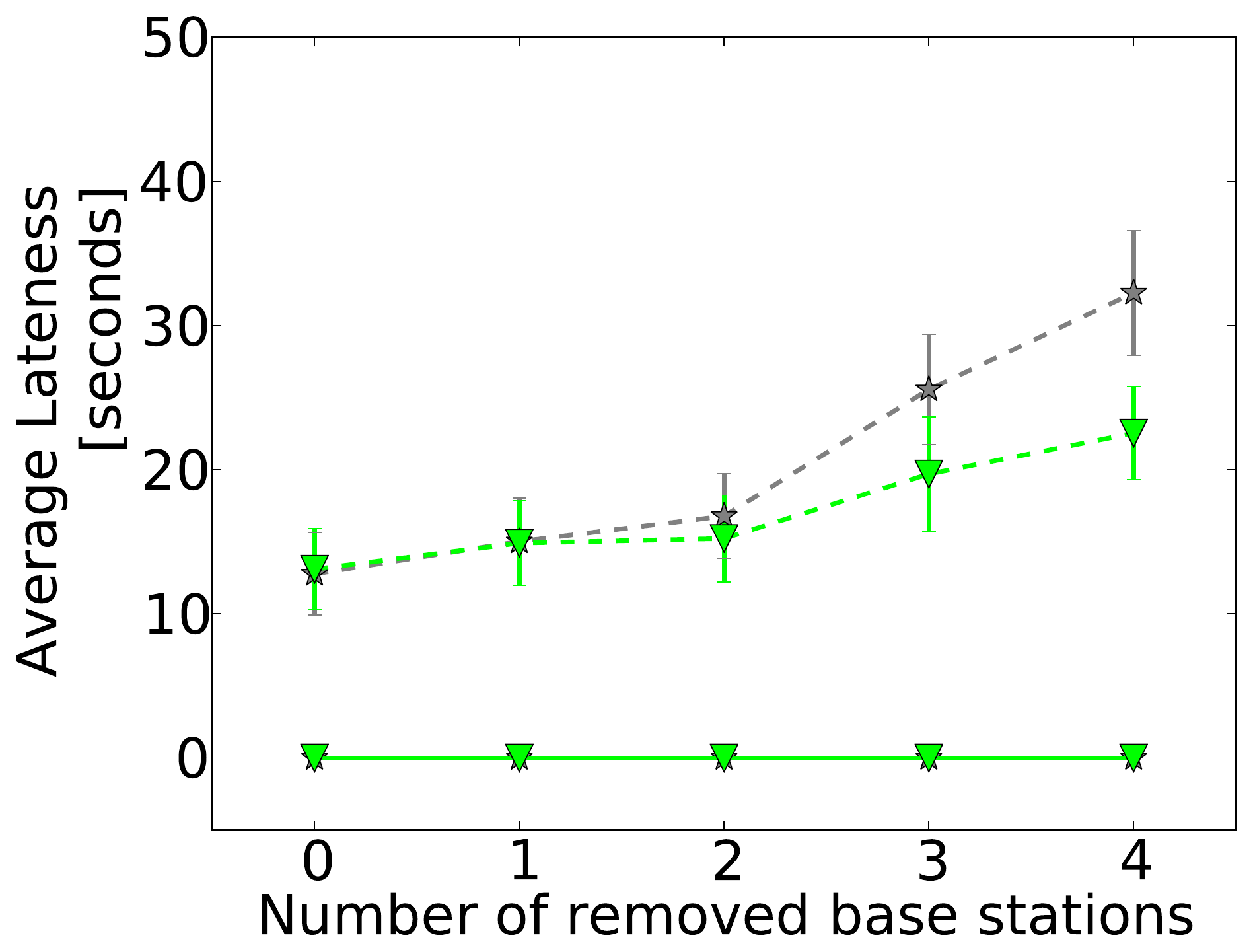}
  \label{fig:tb_lateness}
  }
    %\\
   \subfloat[Average Buffering]{
   \includegraphics[width=0.28\textwidth]{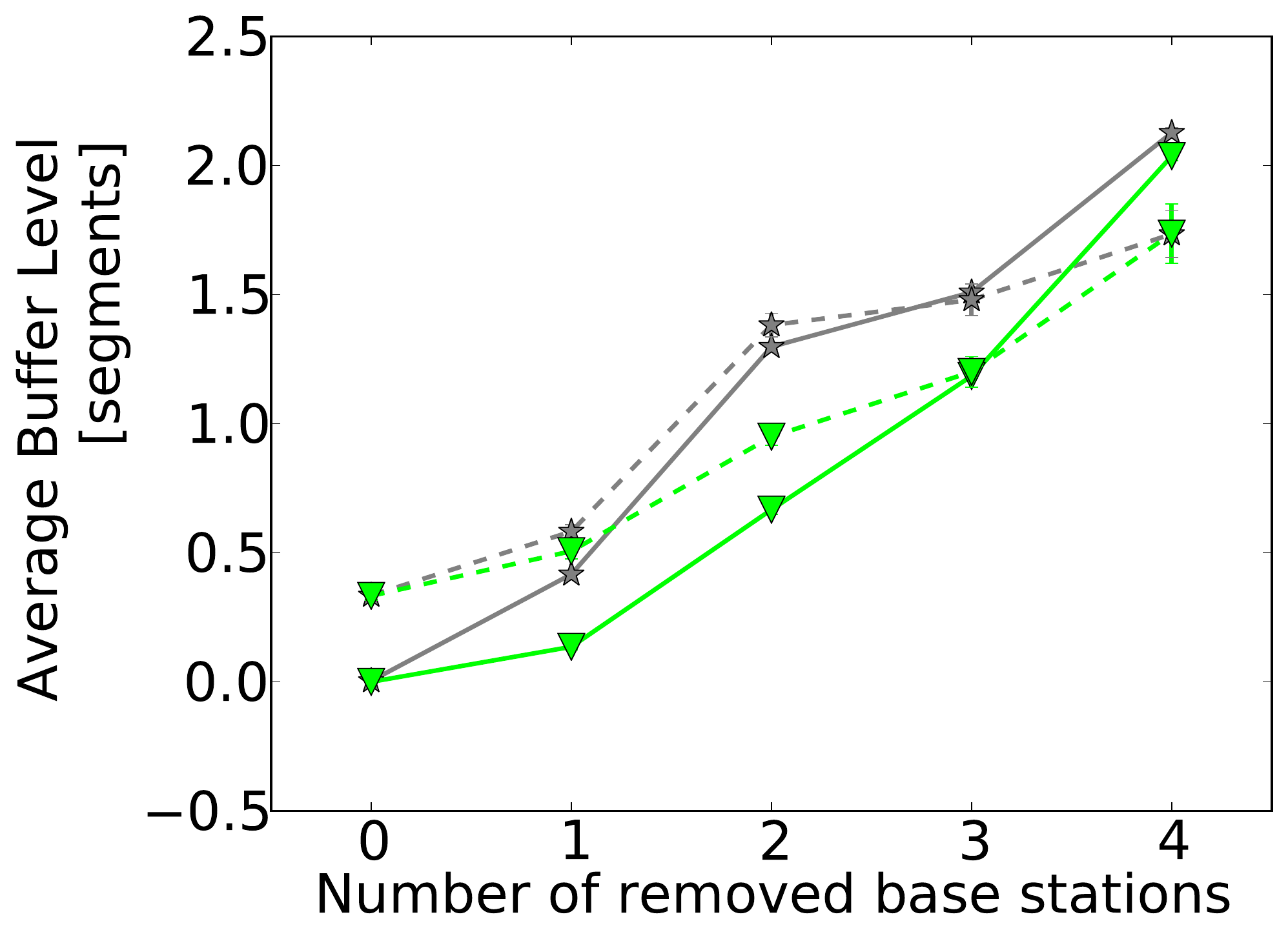}
   \label{fig:tb_buffer}
   }%\\
  \subfloat{
  \includegraphics[width=0.12\textwidth]{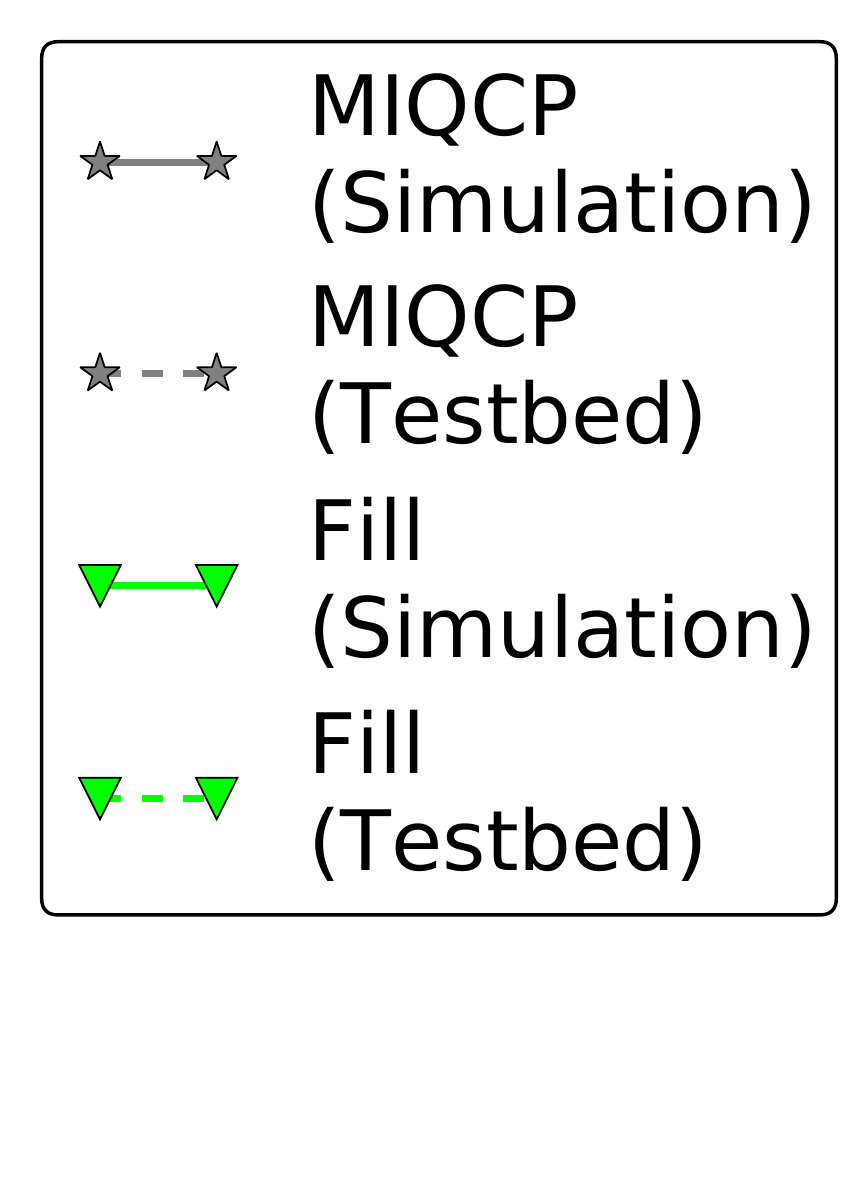}
  \label{fig:tb_legend}
  }
  %\caption{Testbed Measurement Results}
  \caption{Testbed Measurement Results (dashed lines) compared to Simulation Results (solid lines)}
  \label{fig:tb}
  \end{center}
\end{figure*}

For both the simulation and the testbed measurements, we evaluate three different metrics: the average downloaded video quality level in MB per segment, the lateness averaged over all users in seconds and the average buffer fill level
in segments.
All plots are based on multiple simulation or testbed runs and show confidence intervals at 95\% confidence level, small intervals might be covered by the plot markers.

\subsubsection{Simulation Results}

The simulation results for the average video quality are shown in \refFig{fig:sim_quality}.
%The dashed lines indicate the reference value of the high and medium video quality level from \refTab{tab:vquali}
The dashed lines indicate the reference value of the high and medium video quality levels.
MIQCP delivers the overall highest video quality level, which decreases only slightly once more than $10$ base stations
are removed from the scenario. This indicates that MIQCP can fully exploit the available data rate in order to deliver and buffer high quality segments whenever possible.
The \proc{QualityFirst}
algorithm delivers the overall second highest video quality level, which is only slightly less than the one from the MIQCP. This corresponds to the expected behavior of the greedy algorithm.
The \proc{BufferFirst} algorithm exhibits the opposite behavior and delivers the overall lowest video quality level, which also corresponds to the expected behavior. 
The \proc{Fill} algorithm provides the same high video quality level as the MIQCP when only a small number of base stations is removed and enough data rate is available. When more base stations
are removed the delivered video quality level from the \proc{Fill} scheduler decreases, but is still higher compared to the \proc{BufferFirst} algorithm.

\refFig{fig:sim_lateness} shows the results for the average lateness over all users in the simulation. MIQCP and the \proc{Fill} algorithm are able to prevent any lateness.
%The lateness increases when more than $10$ base stations are removed for both the \proc{QualityFirst} and the \proc{BufferFirst} algorithms. Because of the objective to download segments in higher quality levels instead of buffering more segments, the \proc{QualityFirst} algorithm incurs the overall highest lateness.
For both the \proc{QualityFirst} and \proc{BufferFirst} algorithms lateness increases when more than $10$ base stations are removed.
%Because of the objective to download segments in higher quality levels instead of buffering more segments, the \proc{QualityFirst} algorithm incurs the overall highest lateness.
Because of the objective to download segments in higher quality levels instead of buffering more segments, the \proc{QualityFirst} algorithm incurs the highest lateness.

The simulation results for the average buffer fill level are shown in \refFig{fig:sim_buffer}. Both greedy scheduling algorithms always try to fill their buffer up to the maximum buffer level of $3$ segments.
Because the greedy scheduling algorithms have no mechanism to reduce buffer usage, the buffer levels only decrease when there is not enough available data rate to fill the buffer entirely as more base stations
are removed from the scenario. The MIQCP and the \proc{Fill} scheduler are designed to minimize buffer usage where possible, thus both start off with very little buffering and only increase the buffer usage
as more base stations are removed from the scenario. After removing more than $10$ base stations from the scenario the MIQCP uses more buffer space than the \proc{Fill} algorithm. This is caused by the
preference of the MIQCP objective function to download segments with a higher video quality level before minimizing the buffer level.
The\proc{Fill} algorithm, on the other hand, will switch to lower video quality levels before buffering more segments instead.

%\subsubsection{Scheduling Running Times}

% \begin{figure}[htb]
% %\begin{figure}[H]
%   \begin{center}
%   \includegraphics[width=0.28\textwidth]{figures/sim_duration.pdf}
%   \label{fig:sim_durarion}
%   \caption{Scheduling running times; legend in \refFig{fig:sim}}
%   \tix{HK: normiert worauf? legende?}
%   \end{center}
% \end{figure}

% \refFig{fig:sim_durarion} shows the running times for the different scheduling algorithms and the MIQCP on the simulation scenario. All schedulers are executed using a single Intel Xeon X5650 CPU at
% 2.67\,Ghz.
% 
% First of all, running the MIQCP takes around $2.5$ orders of magnitude longer than the greedy and \proc{Fill} algorithms. The running time of the \proc{Fill} algorithm does not change significantly when more
% base stations are removed from the scenario. The running time of the greedy scheduling algorithms slightly decreases as more base stations are removed from the scenarion, because with less available data rate, 
% the greedy algorithms have to make fewer decisions on how to use the available data rate.

\subsubsection{Testbed Measurements}

The plots in \refFig{fig:tb} show a comparison between simulation results with the testbed scenario and the measurements obtained from the testbed. The results from the simulation are plotted with a solid
%line and the testbed measurements with a dashed line, both using the same markers to distinguish between the different schedulers.
line and the testbed measurements with a dashed line, both using the same markers to distinguish between the schedulers.

Ideally, the simulation results and the testbed measurements should be 
identical. 
%Differences in the results can be due to the following effects, which are present in the testbed but not considered in the simulation:
Differences in the results are due to the following effects, which are present in the testbed but not considered in the simulation:
\begin{itemize*}
\item \emph{Continuous time}\\
The simulation is based on a discrete time model with time slots, whereas the testbed runs in real time. In order to compare the simulation and testbed results the measurements are converted to discrete
time. 
This, for example, implies that a segment that is actually downloaded after 61 seconds, but should have been downloaded at or before 60 seconds is treated as equally late as a segment that is downloaded 
after 69 seconds.
%This, for example, implies that a segment that should be dowloaded before 60 seconds have elapsed but is downloaded after 61 seconds is treated as equally late as a segment that is downloaded after
%69 seconds.
\item \emph{Network protocol side effects}\\
The simulation does not consider underlying network protocols for the transport of the HLS segments. In contrast to that the testbed uses real HLS over TCP/IP over 802.11g wireless LAN
with its own wireless resource scheduler.
%We have no influence on whether the data rate limits we use in the calculation of the schedules are actually fully achieved in the testbed.
We are only sure that the data rate limits we use in the calculation of the schedules are not exceeded, but we cannot ensure that they actually fully achieved in the testbed.
Both TCP congestion control and the wireless resource scheduler can influence the actual data rates in the testbed, which result in longer segment downloads, which are then treated as late.
%, which can result in timing side effects.
\item \emph{Video player issues}\\
In case the video player in the testbed has issues while decoding the video, the timing between the downloads from the player and the schedule can be disturbed. For example, if VLC decides to skip
frames from the video the playback runs ahead of the calculated schedule, and subsequent segments are needed for playback before their download was scheduled to be complete.
This can happen because the video player runs on a real Android device and has to share the CPU with the system and background processes.
\end{itemize*}

The measurement results for the average video quality in \refFig{fig:tb_quality} show only little differences between the simulation and testbed. This indicates that our mechanisms for quality selection
work in our testbed implementation as well as expected based on the simulation.

\refFig{fig:tb_lateness} shows the results for the average lateness in the testbed. The measurement results for the greedy schedulers again show only a small difference compared to the simulation, but the 
measurement results for the MIQCP and the \proc{Fill} scheduler show a significantly higher lateness for the testbed. We discovered that this is due to the buffer minimization in these two schedulers: being
forced to use a low buffer level makes the video player more susceptible to the timing side effects we previously described.

The results for the average buffer fill level in \refFig{fig:tb_buffer} again show only a small difference between the simulation results and the testbed measurements.

Taking into account the side effects from the testbed setup, we can sum up that our testbed implementation of the anticipatory scheduling works as forecasted by the simulation results.
This agreement of results between two different and independent evaluation methodologies lends considerable evidence to the utility and feasibility of our proposed anticipatory scheduling scheme.

%%% all lines in one plot
% \begin{figure*}[htb]
% %\begin{figure}[H]
%   \begin{center}
%   \subfloat[Average Quality]{
%   \includegraphics[width=0.28\textwidth]{figures/tb_quality.pdf}
%   \label{fig:tb_quality}
%   }
%   %\\
%   \subfloat[Average Lateness]{
%   \includegraphics[width=0.28\textwidth]{figures/tb_lateness.pdf}
%   \label{fig:tb_lateness}
%   }
%     %\\
%    \subfloat[Average Buffering]{
%    \includegraphics[width=0.28\textwidth]{figures/tb_buffer.pdf}
%    \label{fig:tb_buffer}
%    }%\\
%   \subfloat{
%   \includegraphics[width=0.12\textwidth]{figures/tb_legend.pdf}
%   %\includegraphics[width=0.15\textwidth]{figures/legend2.pdf}
%   \label{fig:tb_legend}
%   }
%   \caption{Testbed Measurement Results}
%   \label{fig:tb}
%   \end{center}
% \end{figure*}
%%% 2 by 2 lines

\section{Conclusion and Future Work}
\label{sec:concl}
We have presented an approach to efficiently exploit knowledge of a user's future wireless data rate for wireless video streaming. Our simulation results and testbed measurements consistently show that adapting buffer and video quality to the anticipated wireless data rate essentially eliminates playback interruptions while maintaining a high video quality level.

Of course, the full benefit of this approach can only be exploited when users request a higher data rate for video streaming than base stations provide. While, in peak times, this already happens today \cite{Akamai}, such a lack of resources will clearly intensify in the near future. Then, our approach will help to utilize the available data rate more efficiently.

% In this paper, we did not elaborate on \emph{how} to anticipate the user's wireless data rate. Recent results on predicting long-term channel states \cite{chen13:localization_ml,Chen2013,Anagnostopoulos2012} show that reliable estimates can be obtained in the order of seconds. This prediction further improves in scenarios with stable trajectories and known radio propagation maps (e.g., highways, railroads). Implementing such anticipation mechanisms, is the focus of our future work.

In this paper we explained how user's wireless data rates can be anticipated, but we did not elaborate on how to implement such a mechanism. Recent results on predicting long-term channel states \cite{chen13:localization_ml,Chen2013,Anagnostopoulos2012} show that reliable estimates can be obtained on the order of seconds, which fits our required time scales. This anticipation can be further improved in scenarios with stable trajectories and known radio propagation maps (e.g., highways, railroads). Implementing such anticipation mechanisms is the focus of our future work.

%We have presented an approach to combine buffer control and quality selection for segmented video streaming with anticipatory knowledge of a mobile user's future data rate. Our simulation and testbed evaluations show a consistent result, indicating that our approach can essentially eliminate playback interruptions while maintaining a high video quality level.

%Of course, the full benefit of this approach can only be exploited when the peak data rate demands from the users for high quality video exceed the full available data rate of the base stations. As this will become reality in the near future, our approach will also help to utilize the available data rate more efficiently.

%So far we did not consider \emph{how} to obtain the anticipatory information on the future data rate of the users. Because we look at coarse grain time frames compared to traditional channel state prediction (segment length in the order of seconds, compared to milliseconds), this information should be easier to obtain. Once we implement a mechanism for this, we will also extend our scheduling algorithms from offline scheduling to online scheduling.

% \ti{HK: diskussionskapitel!!! siehe email
% 
% dichte, volle zellen
% 
% zeitskalen
% 
% weitere fangfragen?}
% 
% \begin{itemize}
% 	%\item on-demand video re-encoding
% 	\item quality / bandwidth
% 	\item real-time
% 	\item prediction
% 	\item testbed
% \end{itemize}

\section{Acknowledgements}
This work was partly supported by Bell Labs, Stuttgart within the research collaboration Smarter Phones And smarter Networks (SPAN).

The research leading to these results has received funding from the European Union's Seventh Framework Programme (FP7/2007-2013) under grant agreement n$^\circ$ 318115.
% We gratefully thank Christoph Schniedermeier, Frederic Beister and Matthias Herlich for their implementation support and for the valuable discussions during our work.

%\balance
%\ti{fix bib entries}
\bibliographystyle{abbrv}
\bibliography{bib}

\balancecolumns
% That's all folks!
\end{document}